\documentclass[iop]{emulateapj}
\usepackage{graphics}
\usepackage{longtable}

\newcommand{\degree}{\hbox{$^\circ$}}

\newcommand{\ltsimeq}{\la}
\newcommand{\gtsimeq}{\ga}

\newcommand{\msun}{M$_{\odot}$}

\newcommand{\HI}{H{\sc i}}
\newcommand{\HII}{H{\sc ii}}

\received{}
\revised{}
\accepted{}
\shortauthors{McQuinn et al.}
\shorttitle{Distance Measurement to Leo~P}

\begin{document}
\title{ALFALFA Discovery of the Nearby Gas-Rich Dwarf Galaxy Leo~P. IV. Distance Measurement from LBT Optical Imaging\protect\footnotemark[*]}
\footnotetext[*]{Based on observations made with the Large Binocular Telescope (LBT) Observatory. The LBT is an international collaboration among institutions in the United States, Italy, and Germany. LBT Corporation partners are: The University of Arizona on behalf of the Arizona university system; Istituto Nazionale di Astrofisica, Italy; LBT Beteiligungsgesellschaft, Germany, representing the Max-Planck Society, the Astrophysical Institute Potsdam, and Heidelberg University; The Ohio State University, and The Research Corporation, on behalf of The University of Minnesota, The University of Notre Dame, and The University of Virginia.}
\author{Kristen~B.~W. McQuinn\altaffilmark{1}, 
Evan D.~Skillman\altaffilmark{1},
Danielle Berg\altaffilmark{1},
John M.~Cannon\altaffilmark{2},
John J.~Salzer\altaffilmark{3},
Elizabeth A.~K.~Adams\altaffilmark{4},
Andrew Dolphin\altaffilmark{5},
Riccardo Giovanelli\altaffilmark{4},
Martha P.~Haynes\altaffilmark{4},
Katherine L.~Rhode\altaffilmark{3}
}

\altaffiltext{1}{Minnesota Institute for Astrophysics, School of Physics and
Astronomy, 116 Church Street, S.E., University of Minnesota,
Minneapolis, MN 55455; \ {\it kmcquinn@astro.umn.edu, skillman@astro.umn.edu, berg@astro.umn.edu}} 
\altaffiltext{2}{Department of Physics and Astronomy, 
Macalester College, 1600 Grand Avenue, Saint Paul, MN 55105;  \ {\it jcannon@macalester.edu}}
\altaffiltext{3}{Department of Astronomy, Indiana University, 727 East 3rd Street, Bloomington, IN 47405; \ {\it rhode@astro.indiana.edu, slaz@astro.indiana.edu}}
\altaffiltext{4}{Center for Radiophysics and Space Research, Space Sciences Building, Cornell University, Ithaca, NY 14853; {\it betsey@astro.cornell.edu, riccardo@astro.cornell.edu, haynes@astro.cornell.edu}}
\altaffiltext{5}{Raytheon Company, 1151 E. Hermans Road, Tucson, AZ 85756}

\begin{abstract}
Leo~P is a low-luminosity dwarf galaxy discovered through the blind \HI\ Arecibo Legacy Fast ALFA (ALFALFA) survey. The \HI\ and follow-up optical observations have shown that Leo~P is a gas-rich dwarf galaxy with both active star formation and an underlying older population, as well as an extremely low oxygen abundance. Here, we measure the distance to Leo~P by applying the tip of the red giant branch (TRGB) distance method to photometry of the resolved stellar population from new Large Binocular Telescope (LBT) V and I band imaging. We measure a distance modulus of 26.19$^{+0.17}_{-0.50}$ mag corresponding to a distance of $1.72^{+0.14}_{-0.40}$ Mpc. Although our photometry reaches 3 magnitudes below the TRGB, the sparseness of the red giant branch (RGB) yields higher uncertainties on the lower limit of the distance. Leo~P is outside the Local Group with a distance and velocity consistent with the local Hubble flow. While located in a very low-density environment, Leo~P lies within $\sim0.5$ Mpc of a loose association of dwarf galaxies which include NGC~3109, Antlia, Sextans~A, and Sextans~B, and 1.1 Mpc away from its next nearest neighbor, Leo~A. Leo~P is one of the lowest metallicity star-forming galaxies known in the nearby universe, comparable in metallicity to I~Zw~18 and DDO~68, but with stellar characteristics similar to dwarf spheriodals (dSphs) in the Local Volume such as Carina, Sextans, and Leo~II. Given its physical properties and isolation, Leo~P may provide an evolutionary link between gas-rich dwarf irregular galaxies and dSphs that have fallen into a Local Group environment and been stripped of their gas.
\end{abstract} 

\keywords{galaxies:\ dwarf -- galaxies:\ distances and redshifts -- galaxies:\ photometry -- galaxies:\ stellar content -- galaxies:\ fundamental parameters -- galaxies:\ evolution}

\section{Introduction\label{intro}}
Over the past decade, significant improvements in observational and data processing capabilities have enabled detailed studies of large samples of dwarf galaxies with resolved stellar populations \citep[e.g.,][]{Tolstoy2009}. While previous generations of results were able to study only the closest Milky Way satellites, resolved studies are now possible to distances outside the Local Volume (D$\sim3$ Mpc) \citep{Dalcanton2009, McQuinn2010, Weisz2011}. The census of nearby dwarf galaxies continues to grow, with the addition of systems of different morphologies and environments \citep[i.e., satellite systems, galaxies on the outskirts of larger systems, and isolated dwarfs; see][and references therein]{McConnachie2012}. The expanding number of nearby dwarf galaxies provides rich and diverse data sets that can be used to disentangle the interconnected evolutionary processes of star formation, chemical evolution, stellar feedback, and the impact of galactic environment. 

The blind \HI\ Arecibo Legacy Fast ALFA survey \citep[ALFALFA;][]{Giovanelli2005, Haynes2011} discovered a low-luminosity star-forming galaxy called Leo~P \citep[AGC~208583;][]{Giovanelli2013}. Follow-up optical observations, including WIYN 3.5m BVR and KPNO 2.1m H$\alpha$ imaging, confirmed the presence of both a young, blue stellar population, including a single \HII\ region, and an underlying older, red stellar population \citep{Rhode2013}. The WIYN 3.5m imaging was not deep enough for a definitive distance determination to Leo~P, but placed the galaxy within the Local Volume, and likely within $\sim2$ Mpc. Optical spectroscopy of the \HII\ region enabled a direct measurement of the auroral [O~III] $\lambda$4363 line, yielding an oxygen abundance of 12 $+$ log(O/H) $=7.17\pm0.04$ \citep{Skillman2013}, $\sim3$\% Z$_{\odot}$ \citep[based on a solar abundance of 12 $+$ log(O/H) $=8.68$;][]{Asplund2009}. These measurements show that Leo~P is the lowest metallicity gas-rich galaxy in the Local Volume. Preliminary reduction of new interferometric \HI\ observations appear to show a small amplitude of rotation in Leo P \citep{Bernstein-Cooper2013}.

Given the oxygen abundance measurement, Leo~P can be classified as an extremely metal deficient \citep[XMD; $12 + $log(O/H) $\leq$7.65; e.g.,][]{Kunth2000} galaxy. The paucity of metals in such a low metallicity dwarf is generally attributed to three different mechanisms: (1) inefficient star formation resulting in large gas mass fractions at the present epoch, (2) the removal of metals through supernova-driven galactic winds from the low potential wells of the galaxies, and/or (3) the dilution of the gas-phase metallicity due to the infall of pristine gas from the outer disk or local extragalactic environment. Given its low luminosity and relatively ordered \HI\ kinematics, the extremely low metallicity of Leo~P is likely attributable to its low average rate of star formation (option 1 above), with the possibility of the removal of some of its metals from a previous galactic wind (option 2). On the other hand, as Leo~P lacks both a high rate of star formation and a disrupted gaseous disk typically associated with interactions, it is unlikely that the low metallicity is due to the infall of pristine gas (option 3). 

Because the galaxy luminosity function (LF) predicts a large density of low-luminosity systems \citep[e.g.,][]{Schechter1976, Binggeli1988}, galaxies with properties similar to those of Leo~P are thought to be very numerous. The properties of such low-luminosity galaxies provide important tests to theories of galaxy formation and evolution. However, in part because of their intrinsically small size, galaxies with these extreme properties that also host luminous, high surface brightness star forming regions that are requisite for chemical abundance studies are rare. Thus, these systems are both difficult to detect and their gas-phase metallicities difficult to measure. It is possible that the discovery of galaxies similar to Leo~P will increase substantially through large blind \HI\ surveys, such as ALFALFA. 

A secure distance measurement to Leo~P is required in order to more fully understand its properties and cosmological context. Thus, we have obtained new ground based V and I band imaging of the stellar populations in Leo P with the Large Binocular Telescope (LBT) which reach $\sim$3 mag below the tip of the red giant branch (TRGB). In \S2 we report the observations and data reduction; in \S3 we describe the distance determination method and uncertainties, and place distance dependent properties of Leo~P on an absolute scale. In \S4 we provide context for Leo~P by comparing it to gas-rich dwarf galaxies with similar metallicity values and to dwarf galaxies with similar physical characteristics. We summarize our findings in \S5. In the Appendix we discuss simulations of low-mass galaxies with sparsely populated red giant branches (RGBs), and the difficulties in identifying the TRGB in these types of systems.

\section{Observations and Photometry of the Resolved Stellar Population\label{obs}}
Deep optical observations in the Bessel V and I band filters were obtained with the LBT at the Mt. Graham Observatory using the Large Binocular Cameras (LBC) on the UT date of 2012, November 23. The LBT uses two 8.4~m diameter primary mirrors with adaptive optics. The 2-mirror design allows for simultaneous imaging of the same field in two filters. Each LBC has a field of view of $\sim23$\arcmin\ $\times\sim23$\arcmin\ with a pixel scale of 0.23\arcsec\ pixel$^{-1}$. 

\begin{figure*}[ht]
\includegraphics[width=\textwidth]{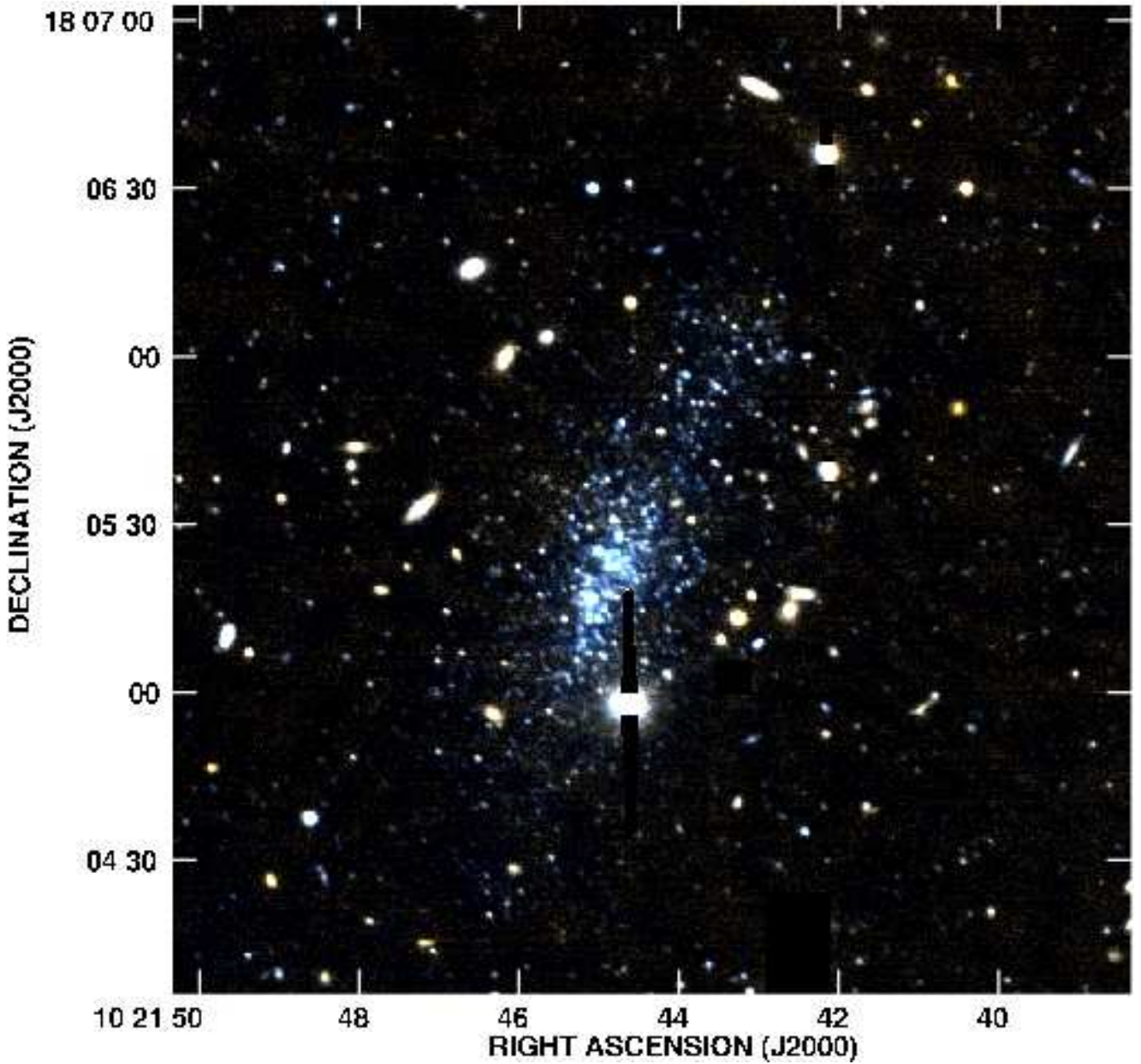}
\caption{Three-color optical image of Leo~P showing both the main star forming complex and underlying older stellar population extending to greater radii. The image was created by combining the V band image (Blue), the average of V and I band images (Green), and the I band image (Red). The observations were obtained from the LBT at the Mt.\ Graham Observatory. The field of view shown here is $\sim1.9$\arcmin$\times\sim2.3$\arcmin, oriented with North-up and East-left. The bleed trail from a saturated foreground star in the image was removed by hand.}
\label{fig:image}
\end{figure*}

The observations consist of 9 images per filter with exposure time of 300 s, dithered with small offsets between every 3 observations to enable cosmic ray and bad pixel rejection. The seeing was $\sim0.7$\arcsec\ with photometric sky conditions. The raw images were bias corrected using combined zero exposures and flat fielded from combined twilight flats taken in the same V and I filters using standard processes in IRAF\footnote[1]{IRAF is distributed by the National Optical Astronomy Observatories, which are operated by the Association of Universities for Research in Astronomy, Inc., under cooperative agreement with the National Science Foundation.}. For each dither position, the three images were averaged to increase the signal to noise and cosmic rays were rejected. Figure~\ref{fig:image} shows a 3-color image combining the V band image (Blue), the average of the V and I band images (Green), and the I band image (Red). 

The observations were designed to reach counts approximating half the full well depth based on average sky background counts in the I band (100 s$^{-1}$ pixel$^{-1}$). However, the background sky was brighter than typical ($\approx$180 c/s/pix), reaching levels of 55,000 counts in the I band images. This high level of background counts, while still in the linear regime, caused problems with the voltage clock of the amplifier reading out the left side of chip 2, which intermittently increased the bias of the CCD. The result was a vertical striping in all I band science images of Leo~P. The stripe pattern was relatively stable across the rows of each image, with an amplitude that varied linearly along the columns. Thus, the background could be characterized with a linear fit in each column (while masking the part of the CCD where Leo~P is located), and subtracted to remove the striping artifact.

The astrometry of the images was determined using SDSS r-band images as a reference. Fifteen stars in the LBT images were individually selected and matched to the same stars in SDSS r-band images of the same field of view. The RA and Dec coordinates from the SDSS image were used with the x and y coordinates from the LBT image as input to the IRAF task \textsc{ccmap} to construct a coordinate transformation for the LBT images. The final astrometry solution is in excellent agreement with that of the WIYN images presented in \citet{Rhode2013}.

Photometry was performed with the DOLPHOT photometry package \citep{Dolphin2000} using an analytically built PSF and a FitSky value of 3. The photometry list was filtered for stars with a signal-to-noise ratio $>4$ and an error flag $<4$. Point sources with high sharpness or crowding values were rejected (i.e., (V$_{sharp}+$I$_{sharp}$)$^2>0.075$; (V$_{crowd}+$I$_{crowd}$)$>0.8$). Artificial star tests were performed to measure the completeness limit of the images using the same photometry package and filtered on the same parameters. 

The photometry was calibrated using a two step process. First, the V band magnitudes were calibrated using the photometry of Leo~P previously obtained on the WIYN 3.5-m telescope at Kitt Peak National Observatory \citep{Rhode2013}. This data set consists of BVR filter photometry calibrated using images of \citet{Landolt1992} standard stars taken before and after the observations. We note that the errors on the photometric calibration coefficients for the WIYN observations were $<$ 0.01 magnitude. The LBT photometry list was matched to this calibrated data using stars with smaller photometric errors and a spatial tolerance of 0.75$\arcsec$. A total of 16 point sources common to the WIYN and LBT images were  matched in the V band magnitude range of $\sim22 - \sim24$ mag. This allowed us to derive a photometric zero-point offset for the LBT data with a formal uncertainty of $\sigma = 0.04$ mag. The calibration showed no dependence on color over the color range of the calibration stars.

Second, as the WIYN data did not include I band imaging, the I band magnitudes from the LBT images were calibrated using the photometry from the SDSS survey. The SDSS r and i filter magnitudes were transformed to Johnson V and I magnitudes using the conversion from Lupton provided on the SDSS website\footnote{http://www.sdss.org/dr5/algorithms/sdssUBVRITransform.html}. This transformation, which includes a color term, was derived by matching the SDSS photometry with published stellar photometry. The SDSS photometry was then matched to the LBT photometry list for the entire LBT field of view with a spatial tolerance of 2$\arcsec$. The matched source list was filtered using additional criteria to ensure the most robust calibration. First, the brightest stars were rejected from the list as they were often saturated in the LBT images. Second, stars fainter than m$_I =$ 22 mag in the SDSS catalog were rejected due to high photometric uncertainties. Third, only point sources were used with SDSS transformed V band magnitudes matching the LBT calibrated V band magnitudes with a tolerance of $\pm0.04$ mag. Finally, the point sources were selected with an SDSS V$-$I color range from $0.5$ mag to 1.6 mag. A total of 6 sources fit these criteria between an I band magnitude range of $\sim20 - \sim22$ yielding a calibration of the LBT I band magnitude with a standard deviation of 0.11 mag. The uncertainty is higher than the uncertainty in V band calibration as the SDSS photometry is limited in the number of stellar sources available for matching. The calibration showed no dependence on color over the color range of the calibration stars. The calibrated photometry was corrected for Galactic absorption using extinction values of $A_V = 0.071$ mag and $A_I = 0.039$ mag from the \citet{Schlafly2011} recalibration of the \citet{Schlegel1998} dust maps.

Spatial cuts were applied to the calibrated photometry. These cuts were determined iteratively by plotting the color magnitude diagram (CMD) of stars in concentric ellipses with the same ellipticity (e $= $0.52) and position angle (PA $= 335\degree$) as Leo~P. In the inner regions, the CMD is dominated by stars from Leo~P, but in the outer regions, it is dominated by background sources. The semi-major and semi-minor axes of the ellipse were increased until the CMDs from larger annuli matched the distribution of point sources from a field region CMD; the final parameter values characterizing the ellipse are listed in Table~\ref{tab:properties}. Table~\ref{tab:catalogue} lists the final photometric catalogue of point sources in Leo~P. 

\begin{figure*}[ht]
\includegraphics[width=\textwidth]{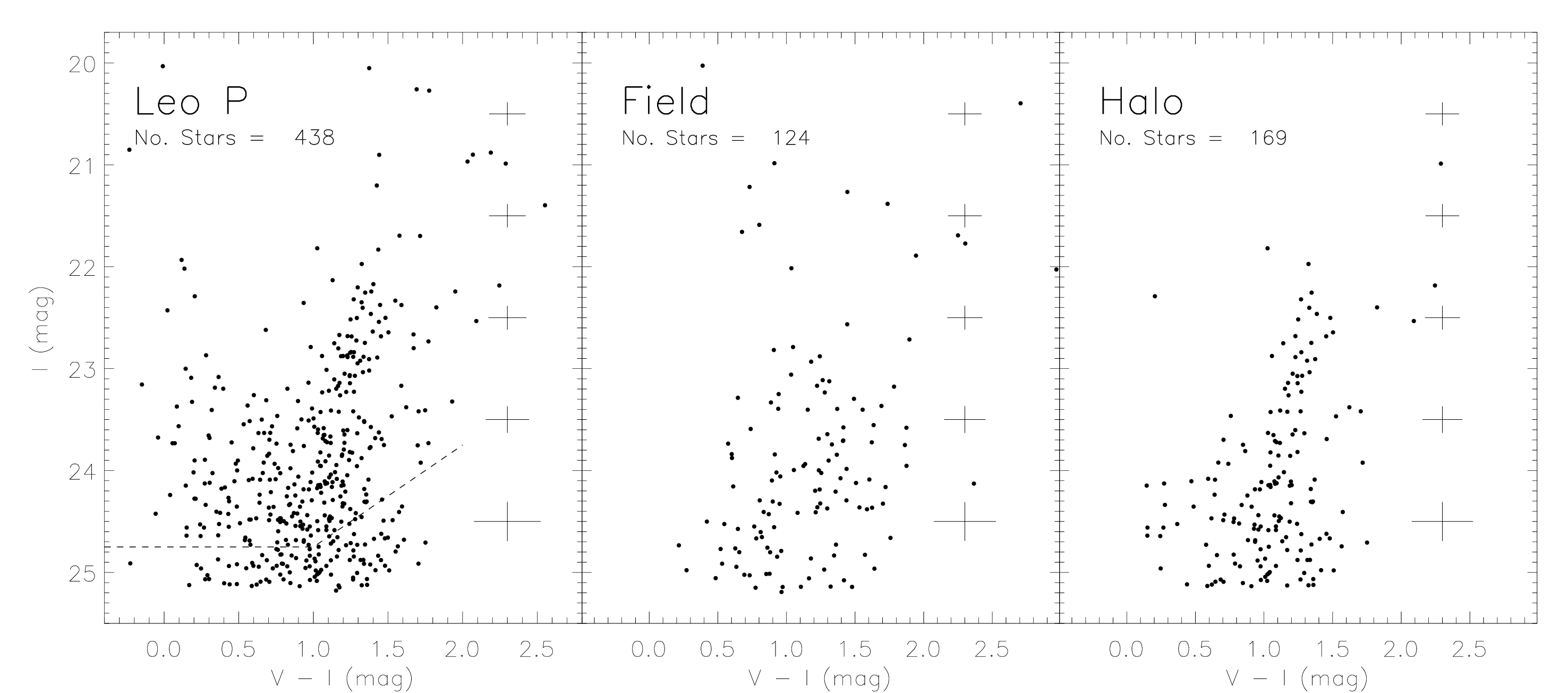}
\caption{\textit{Left panel:} The CMD of Leo~P from V and I band imaging obtained on the LBT. The dashed line marks the 50\% completeness limit determined from artificial star tests. Representative uncertainties per magnitude bin are plotted and include both photometric and calibration uncertainties. The MS and RGB are seen in the CMD. \textit{Middle panel:} CMD of a representative field region of equal area spatially adjacent to Leo~P. The field region lacks the blue MS stars and upper RGB stars seen in the CMD on the left. \textit{Right panel:} CMD of stars from outside the central star-forming region in the stellar halo of Leo~P. The RGB is clearly detected in this CMD with minimal field contamination in the region of smaller areal coverage.}
\label{fig:cmd}
\end{figure*}

The left panel of Figure~\ref{fig:cmd} shows the color-magnitude diagram (CMD) from the final photometry. The dashed line represents the 50\% completeness level determined from the artificial star tests. Representative uncertainties per magnitude are plotted and include both photometric and calibration uncertainties. Both the main sequence (MS) and a sparsely populated RGB are identifiable in the CMD. The middle panel of Figure~\ref{fig:cmd} shows a CMD of a field region of equal size located 1.7$\arcmin$ to the NE of Leo~P. This CMD represents a typical field area in the vicinity of Leo~P and quantifies the possible field contamination. Note there is an absence of stars with V$-$I colors $<0.5$ mag, the approximate color of the MS stars in Leo~P. Conversely, there is a population of contaminating point sources between  V$-$I colors of 0.7 and 1.0 mag. A similar population is seen in the CMD of Leo~P where one would typically expect very few stars between the MS and the RGB at these magnitudes. Given the small angular area and high Galactic latitude of Leo~P, the expected Galactic contamination in this field is low. Using the Besan\c{c}on stellar population synthesis model of the Galaxy \citep[][and references therein]{Robin2003}, less than 10 stars in the photometric range of our data are expected in this field of view. The additional sources in the field CMD are therefore likely background sources, however it cannot be ruled out that some of the sources may be part of an extended stellar disk of Leo~P. Finally, there is a population of stars in the region of the RGB, but very few sources at the brighter magnitudes in the upper RGB. The right panel of Figure~\ref{fig:cmd} shows the CMD populated from stars in the stellar halo of Leo~P (i.e., stars located outside of a central $\sim27$\arcsec$\times\sim14$\arcsec\ ellipse centered on the main star forming region). Field contamination in this CMD is reduced as the areal coverage is smaller. This CMD shows the clear detection of the RGB in the galaxy.

\section{Distance Determinations\label{distance}}
\subsection{TRGB Methodology}
The presence of the RGB in the CMD allows us to use the TRGB method to measure the distance to Leo~P. Briefly, the TRGB distance method is a standard candle approach that arises from the stable and predictable luminosity of low-mass stars just prior to the helium flash \citep{Mould1986, Freedman1988, DaCosta1990}. Photometry in the I band is preferred because the metallicity dependence is reduced compared to that of other optical bands \citep[e.g.,][]{Lee1993}. As a discontinuity in the I band LF identifies the TRGB luminosity, the TRGB detection method requires V and I band observations of resolved stellar populations without significant crowding issues that reach a photometric depth of at least 1 mag below the TRGB luminosity. In addition to a zero-point calibration, a correction for the known metallicity dependence of the TRGB luminosity can be applied. We use the TRGB luminosity calibration from \citet{Rizzi2007} for Johnson-Cousins (JC) filters: 

\begin{equation}
M_I^{JC} = -4.05(\pm0.02) + 0.22(\pm0.01) \cdot [(V-I) - 1.6] \label{eq:trgb}
\end{equation}

\subsection{Measurement of the TRGB in Leo~P}
\begin{figure}[h]
\includegraphics[width=\columnwidth]{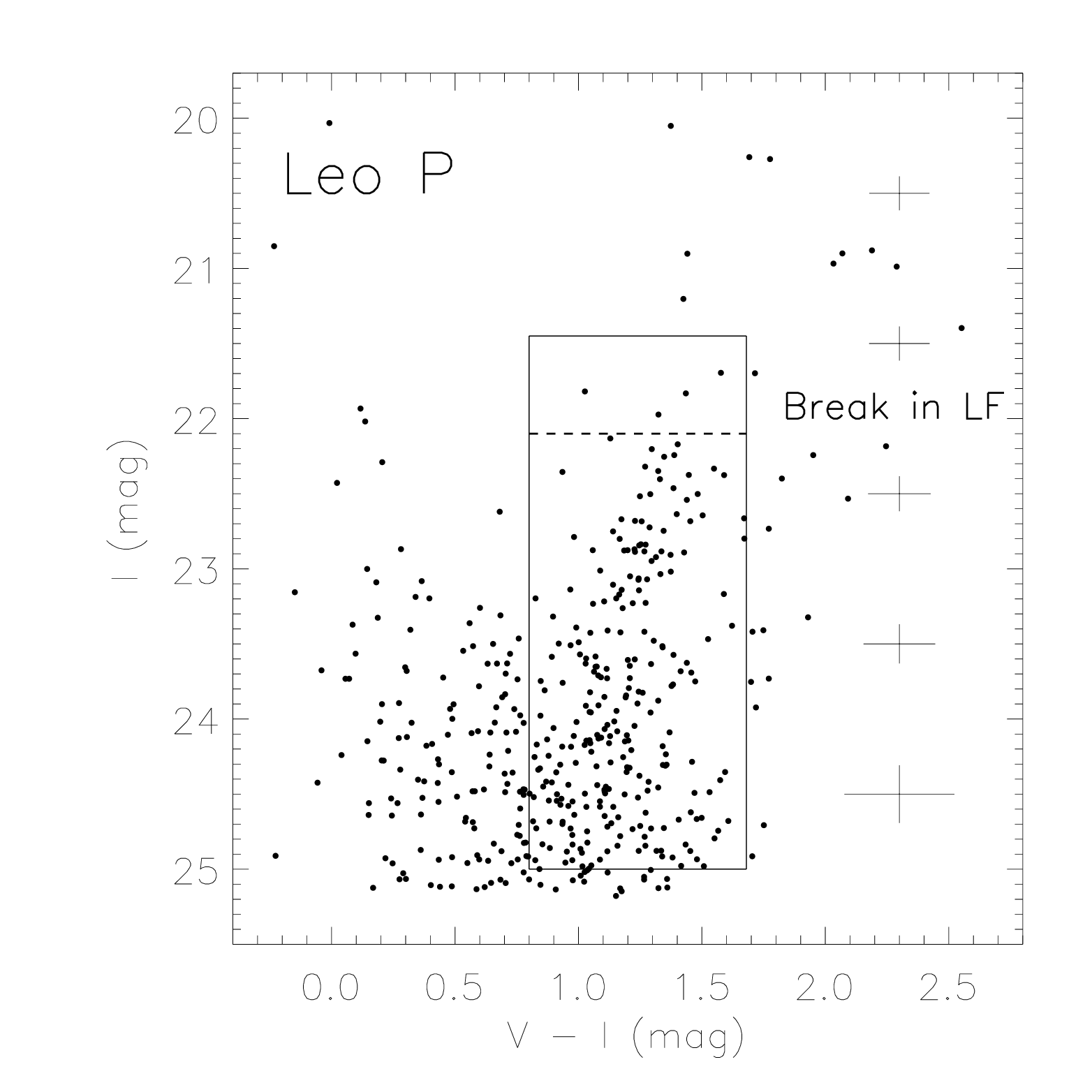}
\caption{The CMD of Leo~P re-plotted with features highlighted. The box encompasses the stars in the RGB region of the CMD used as input to the Sobel filter edge detection. The dashed line at I $= 22.11$ mag demarcates the break in the I band LF identified by a Sobel filter edge detection algorithm shown in Figure~\ref{fig:sobel}. Using a TRGB zero-point calibration and accounting for the metallicity of the TRGB based on the V-I color, this I band magnitude translates to a distance modulus of 26.19 mag and a distance of 1.72 Mpc. Note that given the sparsely populated RGB, the actual TRGB luminosity may be brighter than the identified break in the I band LF. Indeed, there is a grouping of stars brighter than the identified break in the I band LF that may be RGB stars, consistent with simulations of artificial galaxies of comparable stellar mass. If the true TRGB is under-populated and, thus, undetected in CMD, Leo~P may lie closer than our measured distance. See Section~3 and the Appendix for details.}
\label{fig:cmd_trgb}
\end{figure}

The TRGB luminosity was determined by identifying the break in the I band LF. Figure~\ref{fig:cmd_trgb} re-plots the CMD of Leo~P and highlights the region of stars selected for analysis. A Sobel edge detection filter \citep{Lee1993, Sakai1996, Sakai1997} was applied to the I band LF of these stars with a bin width of 0.1 mag. Because the RGB in Leo~P is sparsely populated, small numbers in the I band LF may introduce fluctuations in the Sobel filter response. Thus, we generated 1000 Monte Carlo realizations of the RGB stars based on the photometry of Leo~P by varying the I band luminosities over a range of $\pm4\sigma$ using the I band uncertainties. The fluctuations introduced in the simulated I band LFs are smoothed when averaged over 1000 simulations, thus ensuring a more robust detection of the true break in the I band luminosity function. This is equivalent to the method used by \citet{Sakai1996} in the low-luminosity galaxy Sextans~A.

\begin{figure}[h]
\includegraphics[width=\columnwidth]{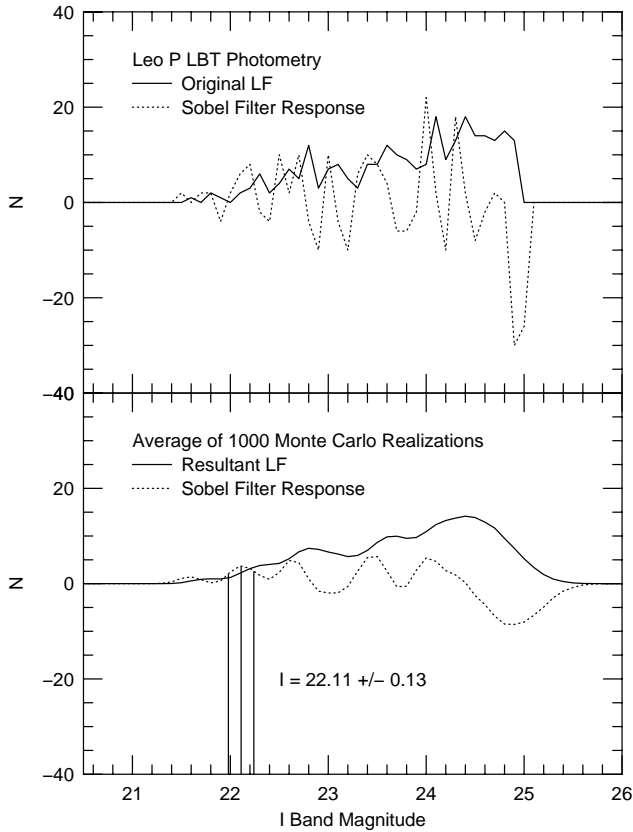}
\caption{\textit{Top panel:} Luminosity function and Sobel filter response from the I band photometry of the region highlighted in Figure~\ref{fig:cmd_trgb}. \textit{Bottom panel:} Luminosity function and Sobel filter response from the average of 1000 Monte Carlo realizations of the I band photometry. The MC realizations reduce the probability of gaps in the I band luminosity function caused by the low number of RGB present in Leo~P being falsely identified as a break in the I band LF. We measure the peak Sobel filter response of the MC realizations to be I $= 22.11$ mag, consistent with the response seen in the data. The width of the Sobel filter response is $\pm0.13$ mag.}
\label{fig:sobel}
\end{figure}

The top panel of Figure~\ref{fig:sobel} shows the I band LF and Sobel filter response from the data; the bottom panel shows the averaged I band LF and the corresponding Sobel filter response from 1000 Monte Carlo simulations. The edge detection in the expected region of the TRGB occurs at m$_I = 22.11$ mag in both the data and the Monte Carlo realizations. In Figure~\ref{fig:cmd_trgb}, this break in the I band LF is marked with a dashed line. Note that there is a weak Sobel filter response at m$_I = 21.60$ mag (based on 4 stars); this will be considered in the discussion of the uncertainties below. In addition, there is a Sobel filter response at m$_I = 23.40$ mag. If this were the TRGB, it would imply that the population of brighter stars in this color range are an extended asymptotic giant branch (AGB) sequence. This is unlikely for two reasons. First, the CMD of halo stars shown in the final panel of Figure~\ref{fig:cmd} includes the majority of the point sources brighter than m$_I = 23.40$ mag in the region of RGB. Since intermediate-age AGB stars are generally not as well-mixed radially in the halos of dwarf galaxies as older-age RGB stars \citep[e.g.,][]{Madore1998}, it is unlikely that these point sources are AGB stars. Second, the average ratio of the number of AGB stars brighter than the TRGB to the number of RGB stars within 2 magnitudes of the TRGB in metal poor galaxies has been shown to be 3.4\% \citep{Girardi2010}. In comparison, if we assume the point sources brighter than m$_I = 23.40$ mag are AGB stars, this ratio would 34\% in Leo~P. As the duration of the AGB phase is $\ltsimeq5$ Myr \citep{Girardi2007}, this large AGB to RGB ratio would imply an extremely unlikely star formation history. It is our intention to study the star formation history of Leo P in detail with Hubble Space Telescope cycle 21 optical imaging (HST-GO-13376, P.I. McQuinn).

Based on the Sobel filter response at m$_I = 22.11$ mag and using an average color of RGB stars of V$-$I $= 1.45$ mag in Equation~\ref{eq:trgb}, the distance modulus of Leo~P is $26.19$ mag -- corresponding to a distance of 1.72 Mpc. This distance determination is in agreement with the estimated distance range of $1.5-2.0$ Mpc given by \citet{Rhode2013}, and within the uncertainties of the estimated baryonic Tully-Fisher distance of $1.3^{+0.9}_{-0.5}$ Mpc given by \citet{Giovanelli2013}. Further consideration of the placement of Leo~P in the baryonic TF is included in Bernstein-Cooper et al. (in prep.), where the rotational dynamics of Leo~P are studied in detail using a suite of interferometric HI spectral line observations.

Typically, the uncertainty on the TRGB distance measurement depends on the uncertainties from the photometry, calibration of the photometry, zero-point calibration of TRGB, and the measurement of TRGB color. In the case of Leo~P, with the exception of the calibration uncertainty on the I band photometry of 0.11 mag, each of these uncertainties is of order 0.05 mag or less. The sparseness of the RGB introduces uncertainties larger than all of these factors. The small number of stars in the CMD dictated that the I band LF be binned in 0.1 mag intervals in order to consistently populate each magnitude bin with a sufficient number of stars. In contrast, for well populated CMDs, the bin width can be as small as 0.01 mag \citep[e.g.,][]{Rizzi2007}. Thus, the width of the Sobel filter response for the LF of Leo~P is also larger. The Sobel filter response on the MC realizations shown in Figure~\ref{fig:sobel} has a half-width half maximum of $\pm0.13$ mag at the peak response of 22.11 mag. For the simpler case of the upper limit of the distance modulus uncertainty, we combine the 0.13 mag width of the Sobel filter response with the uncertainty in the I band calibration of 0.11 mag, which translates to an upper uncertainty on the distance of 0.14 Mpc. 

The lower limit on the distance uncertainty is more difficult to quantify. For sparsely populated RGBs, the measured break in the I band LF may be below the actual TRGB luminosity. For example, in the low mass galaxy Leo~A, the break in the I band LF occurs at $\sim$21.2 mag based on Hubble Space Telescope (HST) photometry \citep{Cole2007}. However, the expected TRGB luminosity is $\sim20.5$ mag based on a distance to Leo~A independently determined using Cepheid variables \citep{Dolphin2002}. Thus, a TRGB distance based on the break in the I band LF from the CMD of Leo~A would have been an upper limit to the actual distance. 

We generated a series of synthetic stellar populations to explore the effect that sparsely populated RGBs have on identifying the TRGB luminosity. A detailed description is presented in the Appendix. Based on the simulations, galaxies with a present-day stellar mass below $\sim10^6$ \msun\ do not fully populate the upper RGB and, thus, the break in the I band LF is fainter than the true TRGB luminosity. For the synthetic galaxy with a stellar mass comparable to Leo~P, the break in the I band LF is a few tenths of a magnitude fainter than the actual TRGB luminosity, with a number of RGB stars in the synthetic CMD that are brighter than the LF break. The CMD of Leo~P shows a similar pattern; there are 5 stars with an I band luminosity up to 0.5 mag brighter than the break in the I band LF. If these stars are RGB stars, then the true TRGB is under-populated and remains undetected in the current data set. As noted above, the Sobel filter shows a weak response to these stars, identifying a change in the I band LF at 21.60 mag. Based on the simulations and the distribution of these point sources brighter than the break in the I band luminosity function, we adopt a conservative lower uncertainty value of 0.5 mag or 0.40 Mpc. Thus, we report the distance to Leo~P with uncertainties to be 1.72$^{+0.14}_{-0.40}$ Mpc. If the true TRGB is indeed undetected, Leo~P may be located at the periphery of the Local Group at a distance of 1.32 Mpc. A second distance determination method based on deeper imaging reaching below the red clump to $M_I~\sim+1$ mag is needed to improve upon this distance measurement.

\subsection{Distance Dependent Quantities and Galactic Environment}
Table~\ref{tab:properties} lists the fundamental properties of Leo~P. The distance from the barycenter of the Local Group (LG) was calculated to be 1.92 Mpc from Equation~\ref{eq:dist_lg}

\begin{equation}
D_{LG}^2 = D^2 + \Delta^2 - 2 \cdot D \cdot \Delta \cdot cos \theta, \label{eq:dist_lg}
\end{equation}

\noindent where D $= 1.72$ Mpc, the distance of the Milky Way from the LG centroid $\Delta = 392$ kpc \citep{McConnachie2012}, and the angle between the direction to Leo~P and that to M31 $\theta = 112^{\circ}$. The total broadband flux of Leo~P was measured with the IRAF task \textsc{polyphot} with an aperture defined from the ellipse parameters provided in Table~\ref{tab:properties}. The absolute V and I magnitudes were determined based on these apparent magnitudes and our distance measurement of 1.72$^{+0.14}_{-0.40}$ Mpc. The present day stellar mass of $5.7\times10^5$ \msun\ was estimated assuming the I band mass to light ratio formalism from \citet{Bell2001}:

\begin{equation}
log~(M/L)_{I} = a_I + b_I \cdot (V - I) \label{eq:ml_ratio}
\end{equation}

\noindent where $a_I = -1.204$, $b_I = +1.347$, and $V-I = 1.08$ mag with an assumed solar luminosity of $M_{I} = 3.33$ mag \citep{Bessel1979, Cox2000}. The stellar mass and luminosity values agree with the range estimated in \citet{Rhode2013}. The surface brightness within the central 20\arcsec\ was measured using the IRAF task \textsc{ellipse} with the elliptical parameters listed in Table~\ref{tab:properties}. The gas mass is based on \HI\ measurements from \citet{Giovanelli2013}. Likewise, the H$\alpha$ luminosity is based on measurements from \citet{Rhode2013}. 

\begin{figure}[h]
\includegraphics[width=\columnwidth]{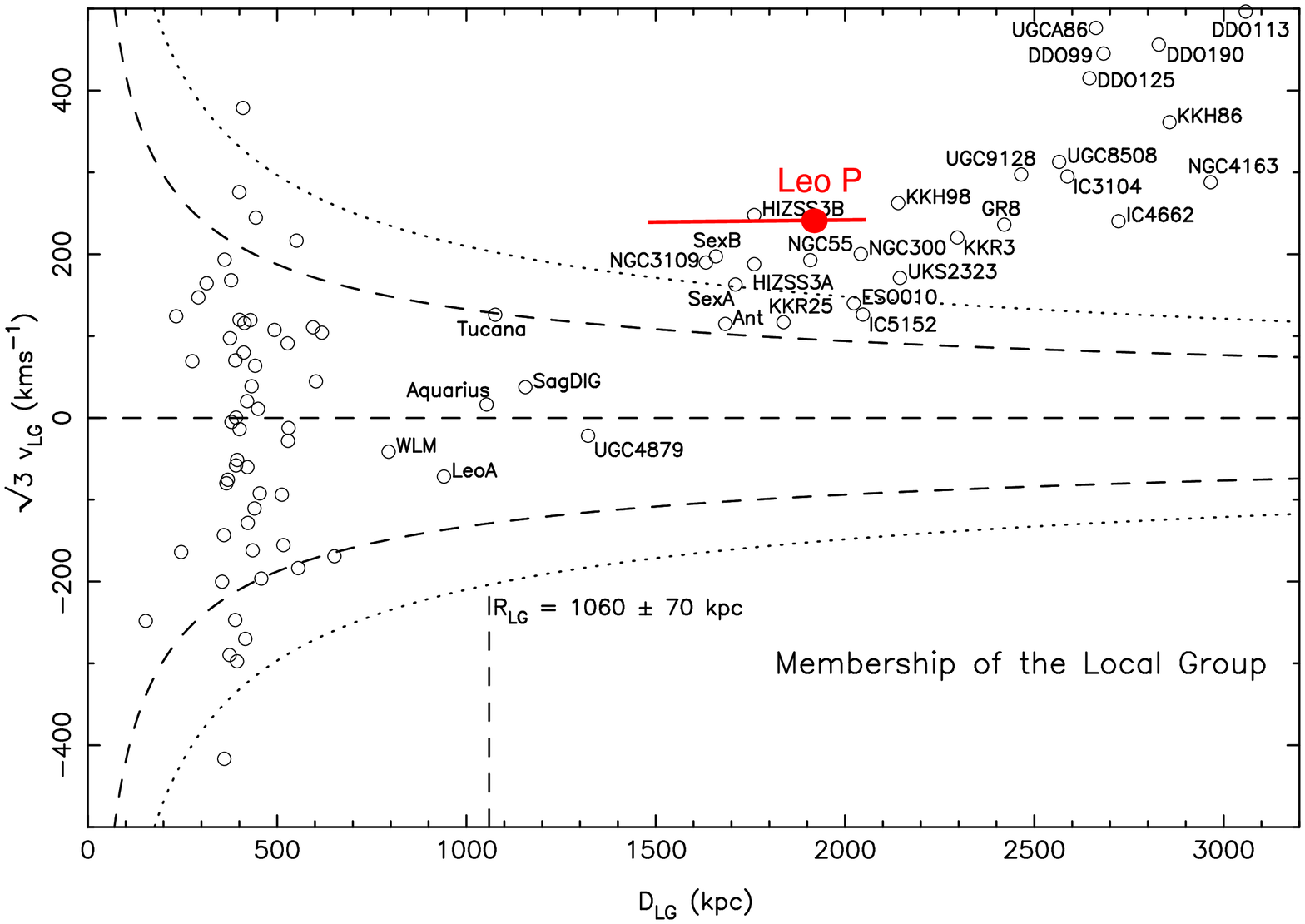}
\caption{A reproduction of a plot from \citet[][Figure~5]{McConnachie2012} showing the distance of both LG and nearby galaxies from the baryonic center of the LG versus the LG-centric velocity of the galaxies. The position of Leo~P is overplotted in red. The LG-centric velocities are multiplied by $\sqrt3$ to account for the unknown tangential velocity components. The dashed and dotted curves indicate the escape velocity from a point mass of $2 \times 10^{12}$ \msun\ and $5 \times 10^{12}$ \msun\ respectively. The unlabeled circles are satellites of either the Galaxy or M31. Leo~P is located outside the virial radius of the LG and is unbound to the LG galaxies. The LG-centric distance and velocity of Leo~P are consistent with the local Hubble flow.}
\label{fig:LG}
\end{figure}

Figure~\ref{fig:LG} is a reproduction of a plot from \citet[][Figure~5]{McConnachie2012} showing the distance of both the LG and nearby galaxies from the baryonic center of the LG versus the LG-centric velocity of the galaxies. The position of Leo~P is overplotted in red. The LG-centric velocities are multiplied by $\sqrt3$ to account for the unknown tangential velocity components. The dashed and dotted curves indicate the escape velocity from a point mass of $2 \times 10^{12}$ \msun\ and $5 \times 10^{12}$ \msun\ respectively. The unlabeled circles are satellites of either the Galaxy or M31. From this Figure, Leo~P is located outside the virial radius of the LG and is not bound to the LG galaxies. Further, the LG-centric velocity and distance of Leo~P are consistent with the local Hubble flow \citep[e.g.,][]{Karachentsev2009, McConnachie2012}. 

\begin{figure}[h]
\includegraphics[width=\columnwidth]{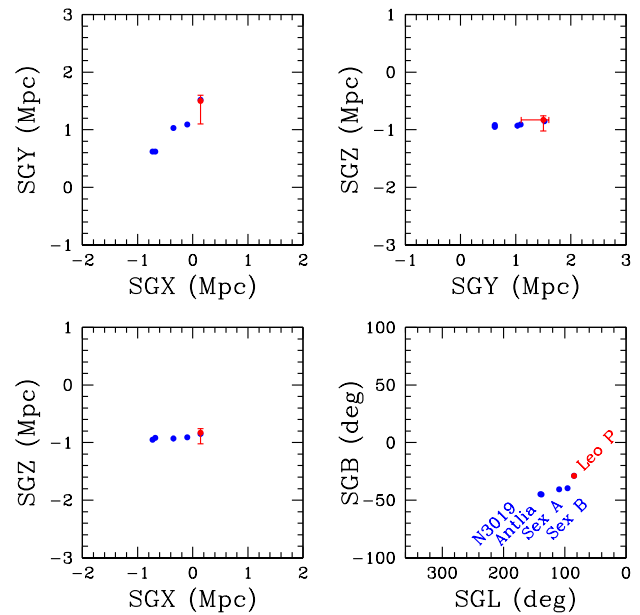}
\caption{The first three panels show plots of the SGX, SGY, and SGZ coordinates of Leo~P in red and the coordinates of the dwarf galaxy association $14+12$ (NGC~3109, Antlia, Sextans~A, Sextans~B) in blue. The final panel shows a plot of the SGL and SGB of the same five galaxies with their names labeled. Uncertainties from the distance measurement to Leo~P are also plotted in the SGY and SGZ coordinates; the error bar in the SGX direction is smaller than the plot point. Leo~P is located $0.47^{+0.14}_{-0.24}$ Mpc from its nearest neighbor, Sextans~B, at the end of a loose association of galaxies enclosing a spherical volume of $\sim0.6$ Mpc$^3$.}
\label{fig:local_environment}
\end{figure}

While Leo~P is located outside of the LG, it is likely part of a small association of dwarf galaxies. Figure~\ref{fig:local_environment} plots the supergalactic (SG) coordinates of Leo~P along with the four galaxies that make up the $14+12$ association, namely NGC~3109, Antlia dwarf, Sextans~A, and Sextans~B \citep{Tully2002}. Because the SG spherical coordinate system has its equator aligned with the major  planar structure in the local universe \citep{deVaucouleurs1953}, SG coordinates help visualize the 3-D spatial distribution of nearby galaxies relative to the planar structure. The first three panels in Figure~\ref{fig:local_environment} plot the SGX, SGY, and SGZ coordinates of these 4 galaxies in blue and those of Leo~P in red. The uncertainties in the SGY and SGZ directions based on our reported distance measurement to Leo~P are shown in the plot; the uncertainties in the SGX direction are smaller than the plot point. The final panel plots the SG latitude (SGL) and longitude (SGB) coordinates of these galaxies with the same colors and labeled with the names of the galaxies. The galaxies previously identified as part of the $14+12$ association are located between $1.25-1.44$ Mpc from the Galaxy within a spherical volume slightly larger than $\sim0.4$ Mpc$^3$ \citep{Tully2006}. Leo~P lies at one end of this association, at a distance of 0.47$^{+0.14}_{-0.24}$ Mpc from Sextans~B. Inclusion of Leo~P in $14+12$ would increase the spherical volume from $\sim0.4$ Mpc$^3$ to just over 0.6 Mpc$^3$, consistent with the volume reported for other dwarf galaxy associations. While the distance to the closest galaxy to Leo~P in the $14+12$ association is $\sim0.5$ Mpc, it is possible that Leo~P is loosely associated with this small grouping of 4 galaxies. This agrees with previous work suggesting that dwarf galaxies exist almost exclusively in some sort of group environment \citep{Tully2006}. 

\begin{figure}[h]
\includegraphics[width=\columnwidth]{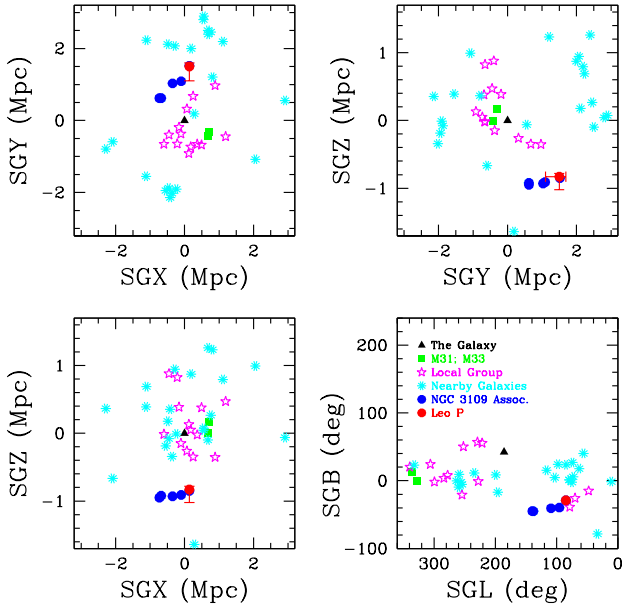}
\caption{The four panels show the same plot configuration of SG coordinates as Figure~\ref{fig:local_environment}, but includes the Galaxy (black triangle), M31 and M33 (green squares), LG galaxies (magenta stars), nearby galaxies outside the LG (cyan asterisks) from \citep{McConnachie2012} as well as the dwarf galaxy association $14+12$ association (blue circles) and Leo~P (red circle). For clarity, satellites of the Galaxy and M31 were omitted. Based on the distribution of galaxies, Leo~P and the nearby loose association of dwarf galaxies are very isolated in the nearby universe.}
\label{fig:greater_environment}
\end{figure}

It is also interesting to consider the greater environment around Leo~P. Figure~\ref{fig:greater_environment} shows the same plots as in Figure~\ref{fig:local_environment}, but adds the Galaxy, M31 and M33, the LG dwarf galaxies, and nearby galaxies from \citet{McConnachie2012} in an expanded field of view. Satellites of the Galaxy and M31 are omitted for clarity. Leo~P and the $14+12$ association are isolated from the LG and other nearby galaxies. The closest system to Leo~P outside of the $14+12$ association is Leo~A at a distance of $\sim1.1$ Mpc. 

From Figures~\ref{fig:local_environment} and \ref{fig:greater_environment}, Leo~P is located at the end of a loose association of dwarf galaxies in an extremely low-density environment. Thus, the evolution of Leo~P has likely been largely unaffected by interactions  from other systems. Leo~P is an excellent candidate for testing theories of galaxy evolution in a truly isolated, low-mass galaxy.

\section{Comparison of Leo~P to Other Dwarf Galaxies \label{discussion}}
It is of interest to compare the properties of Leo~P to other nearby dwarf galaxies. As there are no known systems in the same metallicity regime with similar physical characteristics as Leo~P, we consider three different comparisons. First, we compare Leo~P to dwarf galaxies with comparable gas-phase oxygen abundances (i.e., 12 $+$ log(O/H) $<7.20$).  Second, we compare Leo~P more generally to other XMD galaxies (i.e., 12 $+$ log(O/H) $\leq7.65$). Third, we compare Leo~P to dwarf galaxies with similar properties (i.e., luminosity, stellar mass, and surface brightness values). 

The gas-phase oxygen abundance of Leo~P is among the lowest measured in the XMD classification, comparable to three of the most extreme XMD gas-rich galaxies, namely DDO~68 \citep[12 $+$ log(O/H) $= 7.14\pm0.03$;][]{Pustilnik2005,Izotov2007}, I~Zw~18 \citep[12 $+$ log(O/H) $= 7.17\pm0.04$;][]{Skillman1993, Izotov1999}, and SBS~0335$-$052W \citep[12 $+$ log(O/H) = $7.12\pm0.03$;][]{Izotov2005}. Similarly to Leo~P, these three galaxies have an underlying older stellar population and are not cosmologically young systems \citep[e.g.,][]{Aloisi2007, Pustilnik2004}. However, the inferred mechanism for driving the extremely low metallicities in these galaxies is different from the inferred mechanism for Leo~P. DDO~68, I~Zw~18, and SBS~0335$-$52W are high surface brightness systems discovered through emission-line surveys. All three galaxies have highly disturbed \HI\ morphologies in their gaseous disks, leading to the inference that the very low oxygen abundances are due to infall of low metallicity gas \citep{Ekta2008, Ekta2010a, Ekta2010b}. In I~Zw~18, the neutral gas abundances were recently measured to be a factor of 2 lower than the \HII\ abundances from HST COS data \citep{Lebouteiller2013}, providing further evidence that gas infall is responsible for I~Zw~18's low metallicity. Consistent with an interaction scenario, DDO~68, I~Zw~18, and SBS~0335$-$52W all have high rates of star formation and higher luminosities. As a consequence, these galaxies lie off the well-known luminosity-metallicity (L-Z) relationship for star forming galaxies \citep[e.g.,][and references therein]{Berg2012, Skillman2013}. This correlation between an interaction scenario, higher rates of star formation and luminosities, and lower oxygen abundances is also seen in other XMD galaxies which are slightly more oxygen-rich such as SBS 0335$-$052E \citep{Ekta2009}, SBS 1129$+$576 \citep{Ekta2006}, UGC~772, and J1204$-$0035 \citep{Ekta2008}.

In contrast, the low oxygen abundance in Leo~P does not appear to be driven by a gas infall event. Leo~P is located in a low-density environment and exhibits low levels of star-formation activity. Due to its very low luminosity, it was overlooked in optical surveys and only discovered through a blind \HI\ survey. As previously stated, preliminary reduction of interferometric observations of the neutral hydrogen in Leo~P do not show any discrepant high velocity gas \citep{Bernstein-Cooper2013}, suggesting that Leo~P has not recently undergone a tidal interaction. Thus, while Leo~P has similar oxygen abundances to DDO~68, I~Zw~18, and SBS~0335$-$52W, not only are the overall characteristics quite disparate, the paucity of metals in Leo~P cannot be attributed to the same mechanism. 

Leo~P can be compared with a larger sample of eleven nearby XMD galaxies with slightly higher oxygen abundances from \citet[][see Tables~5 and 7]{Berg2012}. These galaxies are dIrr systems that lie on the L-Z relationship, with the lower luminosity systems being the most metal poor. For galaxies that lie along this relationship, the oxygen abundances are a direct result of normal star-formation activity in the host galaxy, with higher rates of star formation and chemical enrichment correlating with overall galaxy mass. Similarly to these XMD galaxies, Leo~P falls along the L-Z relationship \citep[][see Figure~9]{Skillman2013}, extending the relationship to lower luminosities by nearly two magnitudes. Thus, the extremely low oxygen content in Leo~P is likely a result of an inefficient star formation process at the very low mass end of the galaxy luminosity function. It is also possible that prior galactic winds or outflow events may have been responsible for the partial removal of metals from Leo~P.

Finally, the physical properties of Leo~P can be compared to other nearby dwarf galaxies. Based on the compilation of the properties of galaxies in and around the LG \citep[][and references therein]{McConnachie2012}, the stellar mass, optical luminosity, and central surface brightness of Leo~P are most similar to Carina, Sextans, and Leo~II \citep{Irwin1995}. Table~\ref{tab:parameters} lists values of these physical characteristics for direct comparison. The main difference seen in the comparison is the lower central surface brightness value in Sextans compared to the other three galaxies. On the other hand, the morphology and gas fraction of Leo~P are typical of a dIrr galaxy. Consistent with the morphology-density relationship of galaxies, Leo~P is isolated, whereas the dSphs reside in relatively close proximity to the more massive Milky Way and M~31 galaxies. The possibility that a dIrr can be converted to a dSph through repeated tidal interactions has been successfully modelled by \citet{Mayer2001a,Mayer2001b,Mayer2006}. Although this process is much debated \citep[e.g.,][and references therein]{Grebel2003}, if a dSph progenitor is in the same extremely low mass (and more fragile) regime as Leo~P, the required change is less dramatic \citep{Dekel1986}. Thus, it is possible that Leo~P is representative of a very low-luminosity dwarf galaxy that has not experienced an infall to a LG type environment, retaining its potential to evolve from a low-luminosity, gas-rich dwarf irregular to a low-luminosity, gas-poor dSph.

\section{Conclusions \label{conclusions} }
We have measured the TRGB distance to the newly discovered dwarf galaxy, Leo~P, using optical imaging of resolved stellar populations obtained from ground based observations. We report a distance of $1.72^{+0.14}_{-0.40}$ Mpc. The larger lower limit on the distance reflects the uncertainty in determining the location of the TRGB from the sparsely populated CMD, as discussed in the Appendix. Based on LG-centric distances, velocities, and observed redshift, Leo~P is both outside of and not bound to the LG, and is consistent with the local Hubble flow. Leo~P lies at the end of a previously identified association of 4 other dwarf galaxies: NGC 3109, Antlia, Sextans~A, and Sextans~B \citep[the $14+12$ association,][]{Tully2002}. While at a distance of $\sim0.5$ Mpc from the nearest of these systems, Sextans~B, it is possible that Leo~P is loosely associated with these galaxies. Leo~P and the $14+12$ association are located in an extremely low-density environment; the next nearest galaxy is Leo~A at a distance of $\sim1.1$ Mpc. Thus, the evolution of Leo~P has likely been largely unaffected by interactions with other systems. Leo~P is an excellent candidate for testing theories of galaxy evolution in a truly isolated low-mass galaxy.

Gas-rich, extremely metal-poor galaxies in the mass range of Leo~P are expected to be numerous in the local universe. However, their extremely low luminosities make them difficult to detect. Leo~P, discovered through its \HI\ emission, is the first of such low-luminosity gas-rich star-forming XMD galaxies to be found in the Local Volume.  With an oxygen abundance of 12$+$log(O/H)$= 7.17\pm0.04$ (Skillman et al.\ 2013), Leo P has a metallicity comparable to DDO~68, I~Zw~18, and SBS~0335-52W, but lacks evidence of a significant tidal interaction, and, unlike these other XMD systems, is consistent with the luminosity-metallicity relationship for dwarf star forming galaxies \citep[e.g.,][]{Berg2012, Skillman2013}. It appears that the low oxygen abundance of Leo~P is attributable to its small mass (i.e., inefficient SF, low potential well, or both), and not due to recent infall of metal poor gas as is suggested for other systems with comparable oxygen abundances.

The stellar properties of Leo~P are consistent with the more massive of the dSphs in the Local Group. The stellar mass and luminosity of Leo~P are comparable to Carina, Sextans, and Leo~II, and the central surface brightness of Leo~P is similiar to that of Leo~II and Carina. Yet Leo~P contains a significantly higher gas mass fraction than these dSphs and is located in a very low-density galactic environment. Thus, Leo~P may represent a dSph progenitor that has not experienced an infall into a more densely populated galactic environment.

E.~D.~S. is grateful for partial support from the University of Minnesota. J.~M.~C. is supported by NSF grant AST-1211683. K.~L.~R. is supported by NSF Faculty Early Career Development (CAREER) award AST-0847109. We would like to thank Brent Tully for valuable input. The authors would also like to thank the anonymous referee for helpful and constructive comments. This research made use of NASA's Astrophysical Data System and the NASA/IPAC Extragalactic Database (NED) which is operated by the Jet Propulsion Laboratory, California Institute of Technology, under contract with the National Aeronautics and Space Administration.

{\it Facilities:} \facility{Large Binocular Telescope}

\appendix
\section{Calculation of Lower Uncertainty in a Sparsely Populated RGB \label{appendix}}
As discussed in Section~3, the RGB of Leo~P is more sparsely populated than those seen in more massive dwarf galaxies. Thus, the apparent break in the I band luminosity function may lie below the actual TRGB luminosity in Leo~P. This is seen in another low-mass galaxy, Leo~A, as described above. We tested the sensitivity of populating the TRGB as a function of stellar mass by generating a series of artificial galaxies, and subsequently measuring the break in the I band LF using a Sobel filter. We generated the synthetic stellar populations with the star formation history code MATCH \textsc{fake} tool \citep{Dolphin2002}. Specifically, we used the \textsc{fake} routine with the Padova stellar evolution library of isochrones \citep{Marigo2008}, a Salpeter initial mass function (IMF), and assumed constant values for a star formation rate (SFR), a metal abundance relative to solar value ($-1.50$), a binary fraction, and extinction to generate a series of synthetic stellar populations. Constant SFRs were assumed for each artificial galaxy such that the total stellar mass formed in each simulation ranged from log(M/\msun) $= 5- 7.5$ in increments of 0.5 dex. The present day stellar mass in each galaxy is approximately half this value after accounting for the IMF and the binary fraction. Thus, a total stellar mass of $10^6$ \msun\ approximately corresponds to a present day stellar mass of $5\times10^5$ \msun, comparable to Leo~P. 

\begin{figure}[ht]
\plotone{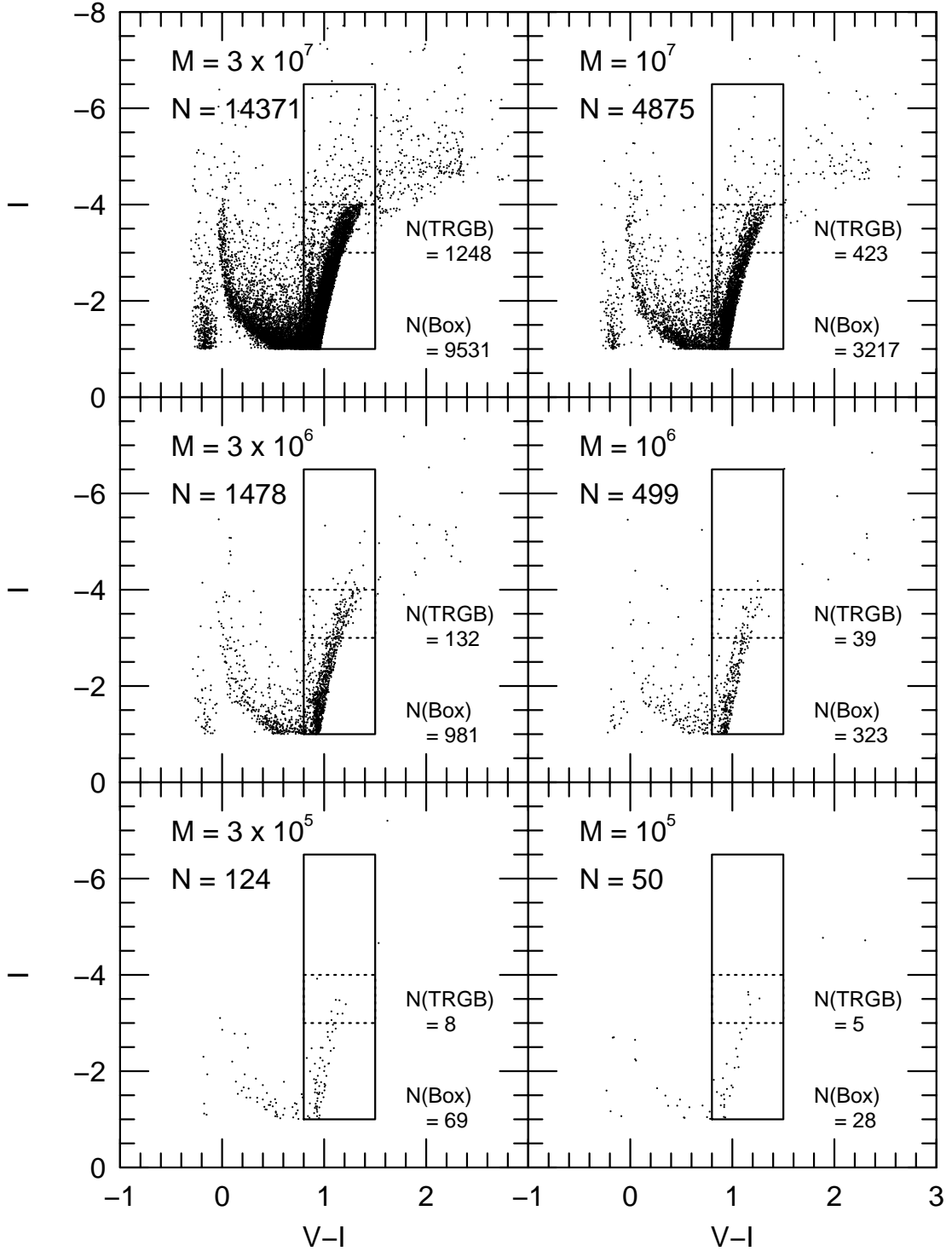}
\caption{Synthetic CMDs generated for six artificial galaxies with total stellar masses ever formed ranging from log(M/\msun) $= 5.0 - 7.5$ in increments of 0.5 dex. Present day stellar masses are approximately half of these values. The boxes encompass the stars in the RGB regions of the CMDs which are used as input to the Sobel filter edge detection algorithm shown in Figure~\ref{fig:synth_sobel}. The top dashed line marks the input TRGB luminosity, the bottom dashed line is one magnitude below this luminosity. The number of stars found in the entire box and in the box reaching 1 magnitude below the TRGB is listed in each panel. Stepping through the panels from the higher to lower mass artificial galaxies, the RGB becomes sparsely populated and the break in the I band LF identifiable by eye does not correspond with the input TRGB luminosity. }
\label{fig:synth_cmds}
\end{figure}

Figure~\ref{fig:synth_cmds} shows the resulting CMDs of six artificial galaxies. The top dashed line represents the input TRGB luminosity of the system (I $= -4.0$ mag), the lower dashed line represents one magnitude below this luminosity. The box in each panel outlines the stars in the RGB region used as input to the Sobel filter edge detection. At the lowest stellar mass, the box encompasses 28 stars in the RGB, with only 5 stars in the magnitude below the TRGB. The artificial galaxy with a total stellar mass formed of $10^6$ \msun\ is the most similar to Leo~P. In this case, there are 39 stars in the magnitude below the TRGB, comparable to a total of 30 stars found in the same region in the CMD of Leo~P. 

\begin{figure}[ht]
\plotone{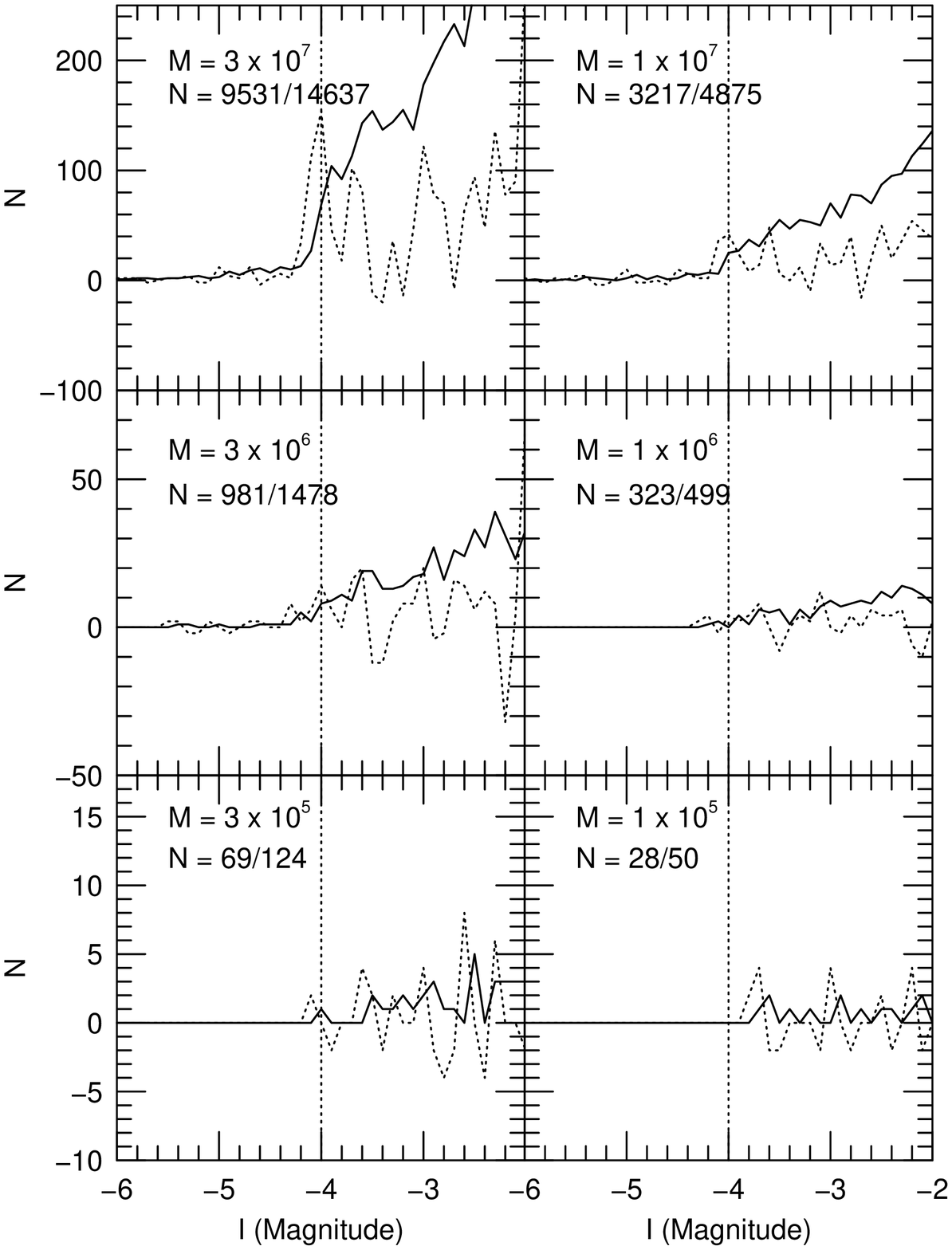}
\caption{The I band LF (solid line) and Sobel filter response (dashed line) for the six artificial galaxies shown in Figure~\ref{fig:synth_cmds}. The vertical dashed line represents the input TRGB luminosity used in the simulations. Stepping through the panels, the peak Sobel filter response corresponding to the break in the I band LF matches the TRGB luminosity for the higher mass systems, but moves to fainter magnitudes for the lower mass systems and is significantly affected by stochastic variability in the lowest mass systems. Thus, a break in the I band LF in a sparsely populated RGB of a galaxy with a present day stellar mass $\ltsimeq10^6$ \msun\ would be fainter than the true TRGB luminosity.}
\label{fig:synth_sobel}
\end{figure}

Figure~\ref{fig:synth_sobel} shows the I band LF of each artificial galaxy (solid line) and the Sobel filter response (dashed line). The vertical dashed line represents the assumed TRGB luminosity used in the simulations (I $= -4.0$ mag). The galaxies are labeled with the total stellar mass formed in the lifetime of the galaxy, equal to approximately twice the present day stellar mass. For the more massive galaxies in the top row (M $\gtsimeq 10^7$ \msun), the Sobel filter response identifying the break in the I band LF corresponds with the TRGB luminosity. For the galaxies in the middle row with a mass between $10^6-10^7$ \msun, the Sobel filter response is variable and affected by one or two individual stars. In these cases, it is difficult to discern which Sobel filter response is representing the break in the LF and the TRGB luminosity. Using the CMDs as a guide, the Sobel filter responses that correspond with the break in the LF are $\sim0.3-0.4$ mag fainter than the TRGB luminosity. This is slightly higher than previously reported uncertainties of $\sim0.2$ mag when measuring the TRGB from a CMD with $\sim50$ stars in the magnitude below the TRGB region using a Sobel filter \citep{Makarov2006}. For the lowest mass galaxies in the bottom row (M $\ltsimeq 3\times 10^5$ \msun), not only is the break in the I band LF identified by the Sobel filter response up to 0.4 mag below the input TRGB luminosity, but there is also a high degree of stochastic variability due to the very small number of stars used in the analysis (i.e., 8 and 5 stars in the magnitude below the TRGB luminosity). Looking more closely at the CMD of Leo~P in Figure~\ref{fig:cmd_trgb}, there is a group of stars in the color range of the RGB up to 0.5 mag brighter than the identified break in the I band LF. Based on similar patterns of stars seen in the simulations, particularly for the artificial galaxy of comparable mass, it is possible that these are RGB stars. If so, then the true TRGB is not well populated in the CMD and the measured break of the I band LF is fainter than the true TRGB luminosity. Given the current data set, it is not possible to discern the true nature of these stars just above the break in the LF, but their presence suggests that Leo~P may be closer than our measured distance. Thus, we adopt a conservative lower uncertainty value of 0.5 mag corresponding to the upper magnitude of this group of point sources, or 0.40 Mpc. Finally, we report the distance with uncertainties to Leo~P to be 1.72$^{+0.14}_{-0.40}$ Mpc.

\begin{table}
\begin{center}
\caption{Fundamental Properties of Leo~P}
\label{tab:properties}
\end{center}
\begin{center}
\begin{tabular}{lr}
\hline 
\hline 
Parameter & Value\\
\hline
R.A. (J2000) 			& 10:21:45.0			\\
Decl. (J2000)			& $+$18:05:01.0			\\
Distance (Mpc)			& 1.72$^{+0.14}_{-0.40}$	\\
D$_{LG}$ (Mpc)			& 1.92$^{+0.13}_{-0.46}$ 	\\
12$+$log(O/H)			& 7.17$\pm0.04$			\\
Semi-major axis ($\arcsec$)	& 70$\pm5$			\\
1-(b/a)				& 0.52				\\
P.A. ($^{\circ}$)		& 335				\\
m$_V$ (mag)			& $16.77\pm0.1$		\\
m$_I$ (mag)			& $15.72\pm0.1$		\\
M$_V$ (mag)			& $-9.41^{+0.14}_{-0.50}$ 	\\
M$_I$ (mag)			& $-10.47^{+0.17}_{-0.50}$	\\
$\mu_{V}$ (mag arcsec$^{-2}$)	& $24.5\pm0.6$	\\
M$_{stars}$ (\msun)		& 5.7$^{+0.4}_{-1.8}\times10^5$ \\
M$_{HI}$ (\msun)		& 9.3$\times10^5$ \\
L$_{H\alpha}$ (erg s$^{-1}$)	& 6.2 $\times10^{36}$ \\
\hline
 \\
\end{tabular}
\end{center}
\tablecomments{Summary of the fundamental properties of Leo~P based on measurements reported in this work and measurements from \citet{Giovanelli2013, Rhode2013, Skillman2013}. Note the elliptical paramters extend to edge of the stellar body detected in the LBT images. The integrated V and I magnitudes were corrected for $A_V = 0.071$ and $A_I = 0.039$ mag of Galactic extintion based on the \citet{Schlafly2011} recalibration of the \citet{Schlegel1998} dust maps. The surface brightness was measured in the central 20\arcsec of Leo~P. See text for details.}

\end{table}

\clearpage
\begin{center}
{\scriptsize
\begin{longtable}{rcccccc}
\multicolumn{7}{c}{\tablename~\thetable{}: LBT Catalogue of Resolved Stars in Leo~P}\\
\\
\hline
        & R.A.          & Decl.         & V$_o$         & $\sigma_V$    & I$_o$         & $\sigma_I$    \\
No.     & (J2000)               & (J2000)       & (mag)         & (mag)         & (mag)         & (mag)         \\
\hline
\hline\hline 
\endfirsthead
\multicolumn{7}{c}{\tablename~\thetable{}: LBT Catalogue of Resolved Stars in Leo~P (Cont'd.)}\\
\\
\hline
        & R.A.          & Decl.         & V$_o$         & $\sigma_V$    & I$_o$         & $\sigma_I$    \\
No.     & (J2000)               & (J2000)       & (mag)         & (mag)         & (mag)         & (mag)         \\
\hline
\hline\hline 
\endhead
\hline\hline
\endlastfoot
1 	 & 10:21:43.27 	 & +18:05:13.36 	 & 20.02 	 & 0.04 	 & 20.03 	 & 0.11 \\
2 	 & 10:21:45.10 	 & +18:05:16.62 	 & 20.62 	 & 0.04 	 & 20.85 	 & 0.11 \\
3 	 & 10:21:44.62 	 & +18:06:09.55 	 & 21.43 	 & 0.04 	 & 20.05 	 & 0.11 \\
4 	 & 10:21:42.62 	 & +18:05:14.59 	 & 21.95 	 & 0.04 	 & 20.26 	 & 0.11 \\
5 	 & 10:21:43.48 	 & +18:05:09.43 	 & 22.05 	 & 0.04 	 & 20.27 	 & 0.11 \\
6 	 & 10:21:45.67 	 & +18:06:03.41 	 & 22.34 	 & 0.04 	 & 20.90 	 & 0.11 \\
7 	 & 10:21:43.10 	 & +18:05:17.26 	 & 23.07 	 & 0.04 	 & 20.88 	 & 0.11 \\
8 	 & 10:21:44.24 	 & +18:05:46.50 	 & 22.97 	 & 0.04 	 & 20.90 	 & 0.11 \\
9 	 & 10:21:45.03 	 & +18:05:24.88 	 & 22.05 	 & 0.04 	 & 21.93 	 & 0.11 \\
10 	 & 10:21:43.85 	 & +18:05:30.70 	 & 23.00 	 & 0.04 	 & 20.97 	 & 0.11 \\
11 	 & 10:21:46.33 	 & +18:04:56.01 	 & 23.28 	 & 0.04 	 & 20.99 	 & 0.11 \\
12 	 & 10:21:44.90 	 & +18:05:22.38 	 & 22.16 	 & 0.04 	 & 22.02 	 & 0.11 \\
13 	 & 10:21:44.83 	 & +18:05:30.87 	 & 22.63 	 & 0.04 	 & 21.20 	 & 0.11 \\
14 	 & 10:21:46.80 	 & +18:05:24.73 	 & 23.95 	 & 0.05 	 & 21.40 	 & 0.11 \\
15 	 & 10:21:44.81 	 & +18:05:22.31 	 & 22.45 	 & 0.04 	 & 22.43 	 & 0.12 \\
16 	 & 10:21:43.92 	 & +18:05:56.59 	 & 22.50 	 & 0.04 	 & 22.29 	 & 0.12 \\
17 	 & 10:21:44.15 	 & +18:05:03.11 	 & 22.84 	 & 0.04 	 & 21.82 	 & 0.11 \\
18 	 & 10:21:42.33 	 & +18:05:46.87 	 & 23.27 	 & 0.04 	 & 21.70 	 & 0.11 \\
19 	 & 10:21:42.43 	 & +18:05:17.38 	 & 23.41 	 & 0.05 	 & 21.70 	 & 0.11 \\
20 	 & 10:21:41.62 	 & +18:05:47.94 	 & 23.27 	 & 0.04 	 & 21.83 	 & 0.11 \\
21 	 & 10:21:44.73 	 & +18:05:48.13 	 & 23.30 	 & 0.04 	 & 21.97 	 & 0.11 \\
22 	 & 10:21:45.13 	 & +18:05:13.76 	 & 23.01 	 & 0.04 	 & 23.16 	 & 0.12 \\
23 	 & 10:21:45.45 	 & +18:05:13.06 	 & 23.26 	 & 0.04 	 & 22.13 	 & 0.11 \\
24 	 & 10:21:45.01 	 & +18:05:13.21 	 & 23.15 	 & 0.04 	 & 22.87 	 & 0.12 \\
25 	 & 10:21:44.76 	 & +18:05:27.17 	 & 23.14 	 & 0.04 	 & 23.00 	 & 0.12 \\
26 	 & 10:21:43.04 	 & +18:05:08.34 	 & 23.29 	 & 0.04 	 & 22.36 	 & 0.12 \\
27 	 & 10:21:44.70 	 & +18:05:34.23 	 & 23.50 	 & 0.05 	 & 22.20 	 & 0.12 \\
28 	 & 10:21:46.92 	 & +18:05:14.33 	 & 23.57 	 & 0.05 	 & 22.17 	 & 0.11 \\
29 	 & 10:21:44.80 	 & +18:05:16.92 	 & 23.82 	 & 0.05 	 & 22.38 	 & 0.12 \\
30 	 & 10:21:42.99 	 & +18:05:08.98 	 & 23.30 	 & 0.04 	 & 22.62 	 & 0.12 \\
31 	 & 10:21:43.39 	 & +18:05:38.07 	 & 23.60 	 & 0.05 	 & 22.25 	 & 0.12 \\
32 	 & 10:21:44.91 	 & +18:05:38.01 	 & 23.63 	 & 0.05 	 & 22.24 	 & 0.12 \\
33 	 & 10:21:45.14 	 & +18:05:20.31 	 & 23.27 	 & 0.04 	 & 23.09 	 & 0.12 \\
34 	 & 10:21:42.92 	 & +18:06:09.47 	 & 24.43 	 & 0.06 	 & 22.18 	 & 0.12 \\
35 	 & 10:21:46.53 	 & +18:04:48.12 	 & 23.59 	 & 0.05 	 & 22.32 	 & 0.12 \\
36 	 & 10:21:45.12 	 & +18:05:29.39 	 & 23.67 	 & 0.05 	 & 22.35 	 & 0.12 \\
37 	 & 10:21:42.56 	 & +18:05:17.56 	 & 24.19 	 & 0.05 	 & 22.24 	 & 0.12 \\
38 	 & 10:21:43.83 	 & +18:05:51.68 	 & 23.73 	 & 0.05 	 & 22.40 	 & 0.12 \\
39 	 & 10:21:42.46 	 & +18:05:42.80 	 & 23.88 	 & 0.05 	 & 22.33 	 & 0.12 \\
40 	 & 10:21:44.01 	 & +18:05:23.81 	 & 23.45 	 & 0.05 	 & 23.08 	 & 0.12 \\
41 	 & 10:21:44.95 	 & +18:05:06.28 	 & 23.46 	 & 0.05 	 & 23.37 	 & 0.12 \\
42 	 & 10:21:41.66 	 & +18:05:50.86 	 & 23.97 	 & 0.05 	 & 22.38 	 & 0.12 \\
43 	 & 10:21:43.56 	 & +18:05:41.07 	 & 23.77 	 & 0.05 	 & 22.52 	 & 0.12 \\
44 	 & 10:21:44.21 	 & +18:05:57.96 	 & 23.85 	 & 0.05 	 & 22.46 	 & 0.12 \\
45 	 & 10:21:45.28 	 & +18:05:21.64 	 & 23.79 	 & 0.05 	 & 22.50 	 & 0.12 \\
46 	 & 10:21:44.54 	 & +18:05:19.90 	 & 23.53 	 & 0.05 	 & 23.19 	 & 0.12 \\
47 	 & 10:21:44.89 	 & +18:05:24.58 	 & 23.51 	 & 0.05 	 & 23.33 	 & 0.12 \\
48 	 & 10:21:42.97 	 & +18:05:38.43 	 & 24.22 	 & 0.05 	 & 22.40 	 & 0.12 \\
49 	 & 10:21:43.01 	 & +18:05:59.13 	 & 23.98 	 & 0.05 	 & 22.50 	 & 0.12 \\
50 	 & 10:21:44.44 	 & +18:05:32.05 	 & 23.59 	 & 0.05 	 & 23.20 	 & 0.12 \\
51 	 & 10:21:47.03 	 & +18:05:05.51 	 & 23.98 	 & 0.05 	 & 22.54 	 & 0.12 \\
52 	 & 10:21:45.64 	 & +18:04:56.54 	 & 23.84 	 & 0.05 	 & 22.67 	 & 0.12 \\
53 	 & 10:21:45.30 	 & +18:05:27.50 	 & 23.77 	 & 0.05 	 & 22.79 	 & 0.12 \\
54 	 & 10:21:44.69 	 & +18:05:23.55 	 & 23.64 	 & 0.05 	 & 23.68 	 & 0.13 \\
55 	 & 10:21:44.19 	 & +18:05:52.48 	 & 23.91 	 & 0.05 	 & 22.68 	 & 0.12 \\
56 	 & 10:21:42.55 	 & +18:05:47.16 	 & 24.62 	 & 0.06 	 & 22.53 	 & 0.12 \\
57 	 & 10:21:44.88 	 & +18:05:13.50 	 & 24.39 	 & 0.06 	 & 23.02 	 & 0.12 \\
58 	 & 10:21:41.62 	 & +18:05:51.61 	 & 23.94 	 & 0.05 	 & 22.68 	 & 0.12 \\
59 	 & 10:21:44.79 	 & +18:05:24.89 	 & 23.66 	 & 0.05 	 & 23.57 	 & 0.13 \\
60 	 & 10:21:43.12 	 & +18:06:01.88 	 & 23.89 	 & 0.05 	 & 22.75 	 & 0.12 \\
61 	 & 10:21:44.34 	 & +18:05:33.37 	 & 24.03 	 & 0.05 	 & 22.64 	 & 0.12 \\
62 	 & 10:21:42.87 	 & +18:05:33.13 	 & 24.15 	 & 0.05 	 & 22.64 	 & 0.12 \\
63 	 & 10:21:45.21 	 & +18:05:09.09 	 & 23.73 	 & 0.05 	 & 23.41 	 & 0.12 \\
64 	 & 10:21:46.38 	 & +18:04:20.14 	 & 24.01 	 & 0.05 	 & 22.72 	 & 0.12 \\
65 	 & 10:21:42.43 	 & +18:05:59.97 	 & 24.14 	 & 0.05 	 & 22.68 	 & 0.12 \\
66 	 & 10:21:45.29 	 & +18:05:05.72 	 & 23.97 	 & 0.05 	 & 22.80 	 & 0.12 \\
67 	 & 10:21:43.54 	 & +18:05:54.52 	 & 23.93 	 & 0.05 	 & 22.88 	 & 0.12 \\
68 	 & 10:21:43.32 	 & +18:06:00.67 	 & 24.09 	 & 0.05 	 & 22.75 	 & 0.12 \\
69 	 & 10:21:46.73 	 & +18:05:06.98 	 & 24.34 	 & 0.06 	 & 22.66 	 & 0.12 \\
70 	 & 10:21:45.01 	 & +18:05:09.79 	 & 23.80 	 & 0.05 	 & 23.73 	 & 0.13 \\
71 	 & 10:21:45.14 	 & +18:05:12.61 	 & 23.86 	 & 0.05 	 & 23.26 	 & 0.12 \\
72 	 & 10:21:46.09 	 & +18:04:28.56 	 & 24.10 	 & 0.05 	 & 22.87 	 & 0.12 \\
73 	 & 10:21:45.18 	 & +18:05:36.87 	 & 24.09 	 & 0.05 	 & 22.84 	 & 0.12 \\
74 	 & 10:21:47.02 	 & +18:04:49.31 	 & 24.09 	 & 0.05 	 & 22.84 	 & 0.12 \\
75 	 & 10:21:45.61 	 & +18:04:30.41 	 & 24.07 	 & 0.05 	 & 22.88 	 & 0.12 \\
76 	 & 10:21:45.36 	 & +18:05:31.21 	 & 24.06 	 & 0.05 	 & 22.88 	 & 0.12 \\
77 	 & 10:21:43.73 	 & +18:05:51.51 	 & 24.11 	 & 0.05 	 & 22.84 	 & 0.12 \\
78 	 & 10:21:44.04 	 & +18:05:33.77 	 & 24.50 	 & 0.06 	 & 22.73 	 & 0.12 \\
79 	 & 10:21:46.03 	 & +18:05:22.45 	 & 24.12 	 & 0.05 	 & 22.89 	 & 0.12 \\
80 	 & 10:21:45.73 	 & +18:05:25.16 	 & 24.15 	 & 0.05 	 & 22.88 	 & 0.12 \\
81 	 & 10:21:44.69 	 & +18:05:21.07 	 & 23.79 	 & 0.05 	 & 23.73 	 & 0.13 \\
82 	 & 10:21:45.12 	 & +18:05:06.92 	 & 23.92 	 & 0.05 	 & 23.36 	 & 0.12 \\
83 	 & 10:21:41.80 	 & +18:05:46.44 	 & 24.47 	 & 0.06 	 & 22.80 	 & 0.12 \\
84 	 & 10:21:44.99 	 & +18:05:06.70 	 & 24.10 	 & 0.05 	 & 23.01 	 & 0.12 \\
85 	 & 10:21:44.99 	 & +18:05:00.23 	 & 24.22 	 & 0.05 	 & 22.88 	 & 0.12 \\
86 	 & 10:21:44.94 	 & +18:05:14.43 	 & 24.32 	 & 0.06 	 & 22.89 	 & 0.12 \\
87 	 & 10:21:44.03 	 & +18:05:18.23 	 & 24.02 	 & 0.05 	 & 23.20 	 & 0.12 \\
88 	 & 10:21:43.44 	 & +18:05:53.54 	 & 24.24 	 & 0.05 	 & 22.92 	 & 0.12 \\
89 	 & 10:21:42.09 	 & +18:05:57.02 	 & 23.99 	 & 0.05 	 & 23.31 	 & 0.12 \\
90 	 & 10:21:46.19 	 & +18:05:05.07 	 & 24.28 	 & 0.05 	 & 22.91 	 & 0.12 \\
91 	 & 10:21:45.75 	 & +18:05:17.34 	 & 24.24 	 & 0.05 	 & 22.95 	 & 0.12 \\
92 	 & 10:21:44.47 	 & +18:05:25.62 	 & 23.95 	 & 0.05 	 & 23.66 	 & 0.13 \\
93 	 & 10:21:42.37 	 & +18:06:00.07 	 & 24.11 	 & 0.05 	 & 23.14 	 & 0.12 \\
94 	 & 10:21:43.89 	 & +18:05:19.35 	 & 23.98 	 & 0.05 	 & 23.68 	 & 0.13 \\
95 	 & 10:21:43.32 	 & +18:06:05.29 	 & 24.26 	 & 0.05 	 & 23.05 	 & 0.12 \\
96 	 & 10:21:44.42 	 & +18:05:26.78 	 & 24.25 	 & 0.05 	 & 23.11 	 & 0.12 \\
97 	 & 10:21:45.33 	 & +18:05:18.55 	 & 24.31 	 & 0.06 	 & 23.06 	 & 0.12 \\
98 	 & 10:21:44.37 	 & +18:05:29.45 	 & 24.09 	 & 0.05 	 & 23.52 	 & 0.13 \\
99 	 & 10:21:44.00 	 & +18:06:01.04 	 & 24.37 	 & 0.06 	 & 23.04 	 & 0.12 \\
100 	 & 10:21:43.45 	 & +18:06:06.72 	 & 24.32 	 & 0.06 	 & 23.07 	 & 0.12 \\
101 	 & 10:21:45.41 	 & +18:05:02.90 	 & 24.08 	 & 0.05 	 & 23.55 	 & 0.13 \\
102 	 & 10:21:43.80 	 & +18:06:06.17 	 & 24.35 	 & 0.06 	 & 23.07 	 & 0.12 \\
103 	 & 10:21:44.59 	 & +18:05:21.40 	 & 24.15 	 & 0.05 	 & 23.50 	 & 0.13 \\
104 	 & 10:21:43.02 	 & +18:05:49.52 	 & 24.32 	 & 0.06 	 & 23.14 	 & 0.12 \\
105 	 & 10:21:44.34 	 & +18:05:36.99 	 & 24.21 	 & 0.05 	 & 23.32 	 & 0.12 \\
106 	 & 10:21:45.27 	 & +18:05:07.84 	 & 24.11 	 & 0.05 	 & 23.90 	 & 0.14 \\
107 	 & 10:21:44.79 	 & +18:05:27.98 	 & 24.34 	 & 0.06 	 & 23.17 	 & 0.12 \\
108 	 & 10:21:44.15 	 & +18:05:11.12 	 & 24.29 	 & 0.05 	 & 23.23 	 & 0.12 \\
109 	 & 10:21:44.40 	 & +18:05:30.51 	 & 24.32 	 & 0.06 	 & 23.22 	 & 0.12 \\
110 	 & 10:21:44.60 	 & +18:06:00.94 	 & 24.39 	 & 0.06 	 & 23.14 	 & 0.12 \\
111 	 & 10:21:42.70 	 & +18:06:00.81 	 & 24.35 	 & 0.06 	 & 23.20 	 & 0.12 \\
112 	 & 10:21:42.73 	 & +18:06:10.64 	 & 24.22 	 & 0.05 	 & 23.46 	 & 0.13 \\
113 	 & 10:21:43.80 	 & +18:05:25.39 	 & 24.18 	 & 0.05 	 & 23.72 	 & 0.13 \\
114 	 & 10:21:45.26 	 & +18:05:29.82 	 & 24.17 	 & 0.05 	 & 23.89 	 & 0.14 \\
115 	 & 10:21:42.10 	 & +18:05:54.67 	 & 24.45 	 & 0.06 	 & 23.23 	 & 0.12 \\
116 	 & 10:21:45.40 	 & +18:04:43.88 	 & 24.44 	 & 0.06 	 & 23.26 	 & 0.12 \\
117 	 & 10:21:44.01 	 & +18:05:53.99 	 & 24.50 	 & 0.06 	 & 23.23 	 & 0.12 \\
118 	 & 10:21:44.04 	 & +18:05:39.77 	 & 24.26 	 & 0.05 	 & 23.63 	 & 0.13 \\
119 	 & 10:21:44.39 	 & +18:05:25.26 	 & 24.21 	 & 0.05 	 & 24.02 	 & 0.15 \\
120 	 & 10:21:45.07 	 & +18:05:38.33 	 & 24.29 	 & 0.05 	 & 23.57 	 & 0.13 \\
121 	 & 10:21:44.73 	 & +18:05:28.15 	 & 24.38 	 & 0.06 	 & 23.39 	 & 0.12 \\
122 	 & 10:21:46.16 	 & +18:04:22.09 	 & 24.30 	 & 0.05 	 & 23.63 	 & 0.13 \\
123 	 & 10:21:42.04 	 & +18:05:35.10 	 & 24.76 	 & 0.07 	 & 23.17 	 & 0.12 \\
124 	 & 10:21:44.60 	 & +18:05:28.65 	 & 24.34 	 & 0.06 	 & 23.63 	 & 0.13 \\
125 	 & 10:21:44.28 	 & +18:05:17.31 	 & 24.28 	 & 0.05 	 & 24.24 	 & 0.16 \\
126 	 & 10:21:44.60 	 & +18:05:26.06 	 & 24.42 	 & 0.06 	 & 23.50 	 & 0.13 \\
127 	 & 10:21:45.90 	 & +18:04:54.70 	 & 24.29 	 & 0.05 	 & 24.15 	 & 0.15 \\
128 	 & 10:21:42.75 	 & +18:05:59.62 	 & 24.47 	 & 0.06 	 & 23.43 	 & 0.12 \\
129 	 & 10:21:46.26 	 & +18:04:54.06 	 & 24.53 	 & 0.06 	 & 23.41 	 & 0.12 \\
130 	 & 10:21:44.75 	 & +18:05:39.11 	 & 24.49 	 & 0.06 	 & 23.49 	 & 0.13 \\
131 	 & 10:21:45.24 	 & +18:05:11.32 	 & 24.38 	 & 0.06 	 & 23.78 	 & 0.14 \\
132 	 & 10:21:43.01 	 & +18:06:05.85 	 & 24.40 	 & 0.06 	 & 23.70 	 & 0.13 \\
133 	 & 10:21:45.38 	 & +18:05:01.86 	 & 24.48 	 & 0.06 	 & 23.51 	 & 0.13 \\
134 	 & 10:21:44.30 	 & +18:06:18.72 	 & 24.48 	 & 0.06 	 & 23.59 	 & 0.13 \\
135 	 & 10:21:44.98 	 & +18:05:37.45 	 & 24.35 	 & 0.06 	 & 24.02 	 & 0.15 \\
136 	 & 10:21:44.01 	 & +18:05:55.20 	 & 24.59 	 & 0.06 	 & 23.42 	 & 0.12 \\
137 	 & 10:21:45.05 	 & +18:05:10.83 	 & 24.37 	 & 0.06 	 & 24.42 	 & 0.18 \\
138 	 & 10:21:43.66 	 & +18:05:28.50 	 & 24.40 	 & 0.06 	 & 23.90 	 & 0.14 \\
139 	 & 10:21:44.30 	 & +18:05:30.88 	 & 24.41 	 & 0.06 	 & 23.93 	 & 0.14 \\
140 	 & 10:21:45.07 	 & +18:05:20.12 	 & 24.48 	 & 0.06 	 & 24.28 	 & 0.16 \\
141 	 & 10:21:44.80 	 & +18:05:45.38 	 & 24.40 	 & 0.06 	 & 24.13 	 & 0.15 \\
142 	 & 10:21:43.59 	 & +18:05:51.33 	 & 24.69 	 & 0.06 	 & 23.42 	 & 0.12 \\
143 	 & 10:21:45.86 	 & +18:05:20.63 	 & 24.43 	 & 0.06 	 & 24.12 	 & 0.15 \\
144 	 & 10:21:44.51 	 & +18:05:26.46 	 & 24.49 	 & 0.06 	 & 23.74 	 & 0.13 \\
145 	 & 10:21:43.82 	 & +18:05:42.17 	 & 24.58 	 & 0.06 	 & 23.57 	 & 0.13 \\
146 	 & 10:21:42.81 	 & +18:06:30.01 	 & 25.25 	 & 0.09 	 & 23.32 	 & 0.12 \\
147 	 & 10:21:44.25 	 & +18:06:19.07 	 & 24.49 	 & 0.06 	 & 24.00 	 & 0.15 \\
148 	 & 10:21:45.99 	 & +18:04:24.37 	 & 24.63 	 & 0.06 	 & 23.60 	 & 0.13 \\
149 	 & 10:21:45.33 	 & +18:05:16.21 	 & 24.54 	 & 0.06 	 & 23.84 	 & 0.14 \\
150 	 & 10:21:44.45 	 & +18:05:20.65 	 & 24.65 	 & 0.06 	 & 23.58 	 & 0.13 \\
151 	 & 10:21:45.68 	 & +18:05:38.30 	 & 25.00 	 & 0.07 	 & 23.38 	 & 0.12 \\
152 	 & 10:21:44.65 	 & +18:05:27.23 	 & 24.78 	 & 0.07 	 & 23.48 	 & 0.13 \\
153 	 & 10:21:44.85 	 & +18:05:33.74 	 & 24.55 	 & 0.06 	 & 23.86 	 & 0.14 \\
154 	 & 10:21:44.15 	 & +18:05:15.58 	 & 24.49 	 & 0.06 	 & 24.28 	 & 0.16 \\
155 	 & 10:21:43.39 	 & +18:05:47.63 	 & 24.59 	 & 0.06 	 & 23.75 	 & 0.14 \\
156 	 & 10:21:43.42 	 & +18:05:16.56 	 & 24.66 	 & 0.06 	 & 23.63 	 & 0.13 \\
157 	 & 10:21:45.42 	 & +18:05:35.74 	 & 24.99 	 & 0.07 	 & 23.47 	 & 0.13 \\
158 	 & 10:21:46.39 	 & +18:04:56.88 	 & 25.12 	 & 0.08 	 & 23.42 	 & 0.12 \\
159 	 & 10:21:43.96 	 & +18:06:07.78 	 & 24.59 	 & 0.06 	 & 23.92 	 & 0.14 \\
160 	 & 10:21:46.56 	 & +18:05:27.73 	 & 25.16 	 & 0.08 	 & 23.41 	 & 0.12 \\
161 	 & 10:21:46.24 	 & +18:05:30.96 	 & 24.86 	 & 0.07 	 & 23.52 	 & 0.13 \\
162 	 & 10:21:42.88 	 & +18:06:02.37 	 & 24.72 	 & 0.06 	 & 23.65 	 & 0.13 \\
163 	 & 10:21:46.05 	 & +18:05:49.46 	 & 24.86 	 & 0.07 	 & 23.52 	 & 0.13 \\
164 	 & 10:21:42.75 	 & +18:05:19.02 	 & 24.81 	 & 0.07 	 & 23.61 	 & 0.13 \\
165 	 & 10:21:44.97 	 & +18:05:08.97 	 & 24.56 	 & 0.06 	 & 24.18 	 & 0.16 \\
166 	 & 10:21:43.70 	 & +18:05:23.20 	 & 24.58 	 & 0.06 	 & 24.11 	 & 0.15 \\
167 	 & 10:21:45.06 	 & +18:05:04.10 	 & 24.73 	 & 0.06 	 & 23.65 	 & 0.13 \\
168 	 & 10:21:45.28 	 & +18:05:34.77 	 & 24.57 	 & 0.06 	 & 24.17 	 & 0.16 \\
169 	 & 10:21:43.70 	 & +18:05:47.87 	 & 24.67 	 & 0.06 	 & 23.81 	 & 0.14 \\
170 	 & 10:21:43.75 	 & +18:05:43.67 	 & 24.70 	 & 0.06 	 & 23.76 	 & 0.14 \\
171 	 & 10:21:45.01 	 & +18:06:12.52 	 & 24.75 	 & 0.07 	 & 23.68 	 & 0.13 \\
172 	 & 10:21:46.04 	 & +18:05:35.46 	 & 24.78 	 & 0.07 	 & 23.67 	 & 0.13 \\
173 	 & 10:21:43.15 	 & +18:05:22.70 	 & 24.83 	 & 0.07 	 & 23.60 	 & 0.13 \\
174 	 & 10:21:46.27 	 & +18:05:21.57 	 & 24.67 	 & 0.06 	 & 23.93 	 & 0.14 \\
175 	 & 10:21:44.78 	 & +18:05:47.00 	 & 24.61 	 & 0.06 	 & 24.34 	 & 0.17 \\
176 	 & 10:21:43.85 	 & +18:05:46.24 	 & 24.85 	 & 0.07 	 & 23.65 	 & 0.13 \\
177 	 & 10:21:43.58 	 & +18:05:55.18 	 & 24.79 	 & 0.07 	 & 23.71 	 & 0.13 \\
178 	 & 10:21:46.76 	 & +18:05:23.72 	 & 24.96 	 & 0.07 	 & 23.57 	 & 0.13 \\
179 	 & 10:21:46.86 	 & +18:05:01.15 	 & 24.66 	 & 0.06 	 & 24.09 	 & 0.15 \\
180 	 & 10:21:46.44 	 & +18:04:39.78 	 & 24.81 	 & 0.07 	 & 23.72 	 & 0.13 \\
181 	 & 10:21:42.92 	 & +18:05:34.72 	 & 24.67 	 & 0.06 	 & 24.08 	 & 0.15 \\
182 	 & 10:21:44.18 	 & +18:05:18.82 	 & 24.68 	 & 0.06 	 & 24.02 	 & 0.15 \\
183 	 & 10:21:45.31 	 & +18:04:45.73 	 & 24.93 	 & 0.07 	 & 23.63 	 & 0.13 \\
184 	 & 10:21:46.39 	 & +18:04:53.92 	 & 24.84 	 & 0.07 	 & 23.73 	 & 0.13 \\
185 	 & 10:21:46.42 	 & +18:04:27.19 	 & 24.74 	 & 0.07 	 & 23.98 	 & 0.14 \\
186 	 & 10:21:44.65 	 & +18:05:45.39 	 & 24.70 	 & 0.06 	 & 24.27 	 & 0.16 \\
187 	 & 10:21:44.01 	 & +18:04:35.66 	 & 25.07 	 & 0.08 	 & 23.63 	 & 0.13 \\
188 	 & 10:21:46.13 	 & +18:04:48.98 	 & 24.73 	 & 0.07 	 & 24.09 	 & 0.15 \\
189 	 & 10:21:44.15 	 & +18:05:27.16 	 & 24.68 	 & 0.06 	 & 24.91 	 & 0.24 \\
190 	 & 10:21:43.93 	 & +18:05:22.24 	 & 24.94 	 & 0.07 	 & 23.73 	 & 0.13 \\
191 	 & 10:21:43.05 	 & +18:06:15.29 	 & 24.87 	 & 0.07 	 & 23.82 	 & 0.14 \\
192 	 & 10:21:43.35 	 & +18:06:17.13 	 & 24.71 	 & 0.06 	 & 24.56 	 & 0.19 \\
193 	 & 10:21:46.02 	 & +18:05:43.58 	 & 24.80 	 & 0.07 	 & 24.09 	 & 0.15 \\
194 	 & 10:21:44.33 	 & +18:05:21.01 	 & 24.82 	 & 0.07 	 & 23.98 	 & 0.14 \\
195 	 & 10:21:44.23 	 & +18:05:26.55 	 & 24.74 	 & 0.07 	 & 24.30 	 & 0.17 \\
196 	 & 10:21:43.22 	 & +18:05:07.38 	 & 24.80 	 & 0.07 	 & 24.03 	 & 0.15 \\
197 	 & 10:21:44.09 	 & +18:05:33.43 	 & 25.10 	 & 0.08 	 & 23.68 	 & 0.13 \\
198 	 & 10:21:45.53 	 & +18:05:00.53 	 & 24.75 	 & 0.07 	 & 24.40 	 & 0.17 \\
199 	 & 10:21:44.63 	 & +18:05:36.41 	 & 25.00 	 & 0.07 	 & 23.79 	 & 0.14 \\
200 	 & 10:21:45.28 	 & +18:05:42.87 	 & 24.96 	 & 0.07 	 & 23.85 	 & 0.14 \\
201 	 & 10:21:43.15 	 & +18:05:39.20 	 & 25.15 	 & 0.08 	 & 23.69 	 & 0.13 \\
202 	 & 10:21:43.97 	 & +18:05:19.29 	 & 24.83 	 & 0.07 	 & 24.09 	 & 0.15 \\
203 	 & 10:21:44.99 	 & +18:05:33.68 	 & 24.77 	 & 0.07 	 & 24.53 	 & 0.19 \\
204 	 & 10:21:41.74 	 & +18:05:50.41 	 & 25.16 	 & 0.08 	 & 23.78 	 & 0.14 \\
205 	 & 10:21:44.33 	 & +18:05:22.63 	 & 24.79 	 & 0.07 	 & 24.41 	 & 0.18 \\
206 	 & 10:21:43.97 	 & +18:05:37.77 	 & 24.94 	 & 0.07 	 & 23.91 	 & 0.14 \\
207 	 & 10:21:43.71 	 & +18:05:21.69 	 & 24.79 	 & 0.07 	 & 24.64 	 & 0.20 \\
208 	 & 10:21:44.28 	 & +18:05:37.87 	 & 25.04 	 & 0.08 	 & 23.84 	 & 0.14 \\
209 	 & 10:21:44.00 	 & +18:05:47.48 	 & 25.06 	 & 0.08 	 & 23.82 	 & 0.14 \\
210 	 & 10:21:47.02 	 & +18:05:07.97 	 & 25.15 	 & 0.08 	 & 23.77 	 & 0.14 \\
211 	 & 10:21:43.52 	 & +18:05:40.50 	 & 25.00 	 & 0.07 	 & 23.95 	 & 0.14 \\
212 	 & 10:21:46.37 	 & +18:04:44.03 	 & 24.84 	 & 0.07 	 & 24.35 	 & 0.17 \\
213 	 & 10:21:45.67 	 & +18:05:04.05 	 & 24.99 	 & 0.07 	 & 23.91 	 & 0.14 \\
214 	 & 10:21:45.23 	 & +18:05:02.14 	 & 25.09 	 & 0.08 	 & 23.82 	 & 0.14 \\
215 	 & 10:21:43.32 	 & +18:06:03.03 	 & 24.88 	 & 0.07 	 & 24.24 	 & 0.16 \\
216 	 & 10:21:43.39 	 & +18:05:58.94 	 & 25.05 	 & 0.08 	 & 23.86 	 & 0.14 \\
217 	 & 10:21:45.62 	 & +18:05:36.64 	 & 24.83 	 & 0.07 	 & 24.56 	 & 0.19 \\
218 	 & 10:21:46.08 	 & +18:04:27.59 	 & 25.22 	 & 0.09 	 & 23.75 	 & 0.14 \\
219 	 & 10:21:43.86 	 & +18:05:44.90 	 & 24.96 	 & 0.07 	 & 24.06 	 & 0.15 \\
220 	 & 10:21:41.85 	 & +18:05:37.01 	 & 25.50 	 & 0.10 	 & 23.73 	 & 0.13 \\
221 	 & 10:21:45.90 	 & +18:05:06.68 	 & 25.01 	 & 0.07 	 & 23.96 	 & 0.14 \\
222 	 & 10:21:43.79 	 & +18:06:21.69 	 & 24.86 	 & 0.07 	 & 24.43 	 & 0.18 \\
223 	 & 10:21:46.45 	 & +18:05:34.72 	 & 25.01 	 & 0.07 	 & 24.02 	 & 0.15 \\
224 	 & 10:21:42.90 	 & +18:05:26.75 	 & 24.93 	 & 0.07 	 & 24.21 	 & 0.16 \\
225 	 & 10:21:42.09 	 & +18:06:11.16 	 & 25.14 	 & 0.08 	 & 23.90 	 & 0.14 \\
226 	 & 10:21:42.15 	 & +18:05:44.76 	 & 25.45 	 & 0.10 	 & 23.75 	 & 0.14 \\
227 	 & 10:21:45.37 	 & +18:05:47.70 	 & 24.89 	 & 0.07 	 & 24.64 	 & 0.20 \\
228 	 & 10:21:44.74 	 & +18:04:29.56 	 & 25.10 	 & 0.08 	 & 23.95 	 & 0.14 \\
229 	 & 10:21:45.48 	 & +18:05:46.16 	 & 24.89 	 & 0.07 	 & 24.52 	 & 0.19 \\
230 	 & 10:21:44.61 	 & +18:05:30.45 	 & 25.20 	 & 0.09 	 & 23.88 	 & 0.14 \\
231 	 & 10:21:47.04 	 & +18:05:00.04 	 & 25.01 	 & 0.07 	 & 24.14 	 & 0.15 \\
232 	 & 10:21:45.45 	 & +18:05:18.15 	 & 24.95 	 & 0.07 	 & 24.32 	 & 0.17 \\
233 	 & 10:21:44.97 	 & +18:04:39.64 	 & 25.00 	 & 0.07 	 & 24.17 	 & 0.16 \\
234 	 & 10:21:46.15 	 & +18:04:59.74 	 & 25.09 	 & 0.08 	 & 24.11 	 & 0.15 \\
235 	 & 10:21:43.43 	 & +18:06:00.01 	 & 25.16 	 & 0.08 	 & 24.02 	 & 0.15 \\
236 	 & 10:21:43.70 	 & +18:05:36.54 	 & 25.07 	 & 0.08 	 & 24.25 	 & 0.16 \\
237 	 & 10:21:46.41 	 & +18:04:24.12 	 & 25.16 	 & 0.08 	 & 24.04 	 & 0.15 \\
238 	 & 10:21:44.41 	 & +18:06:12.72 	 & 25.25 	 & 0.09 	 & 23.96 	 & 0.14 \\
239 	 & 10:21:45.64 	 & +18:05:17.53 	 & 24.98 	 & 0.07 	 & 24.55 	 & 0.19 \\
240 	 & 10:21:44.44 	 & +18:05:47.11 	 & 25.12 	 & 0.08 	 & 24.18 	 & 0.16 \\
241 	 & 10:21:43.82 	 & +18:05:19.96 	 & 25.00 	 & 0.07 	 & 24.64 	 & 0.20 \\
242 	 & 10:21:45.99 	 & +18:04:59.56 	 & 25.18 	 & 0.08 	 & 24.11 	 & 0.15 \\
243 	 & 10:21:43.80 	 & +18:06:12.35 	 & 25.17 	 & 0.08 	 & 24.07 	 & 0.15 \\
244 	 & 10:21:45.80 	 & +18:05:12.49 	 & 25.07 	 & 0.08 	 & 24.36 	 & 0.17 \\
245 	 & 10:21:43.63 	 & +18:05:14.19 	 & 25.16 	 & 0.08 	 & 24.18 	 & 0.16 \\
246 	 & 10:21:43.67 	 & +18:05:31.43 	 & 25.02 	 & 0.08 	 & 24.52 	 & 0.19 \\
247 	 & 10:21:45.77 	 & +18:04:38.64 	 & 25.18 	 & 0.08 	 & 24.14 	 & 0.15 \\
248 	 & 10:21:46.41 	 & +18:05:30.97 	 & 25.05 	 & 0.08 	 & 24.48 	 & 0.18 \\
249 	 & 10:21:45.35 	 & +18:04:54.31 	 & 25.21 	 & 0.09 	 & 24.16 	 & 0.15 \\
250 	 & 10:21:46.35 	 & +18:05:21.93 	 & 25.12 	 & 0.08 	 & 24.25 	 & 0.16 \\
251 	 & 10:21:45.16 	 & +18:05:32.13 	 & 25.27 	 & 0.09 	 & 24.05 	 & 0.15 \\
252 	 & 10:21:43.86 	 & +18:05:27.01 	 & 25.09 	 & 0.08 	 & 24.36 	 & 0.17 \\
253 	 & 10:21:45.51 	 & +18:05:48.16 	 & 25.21 	 & 0.09 	 & 24.13 	 & 0.15 \\
254 	 & 10:21:45.38 	 & +18:05:28.69 	 & 25.06 	 & 0.08 	 & 24.48 	 & 0.18 \\
255 	 & 10:21:45.62 	 & +18:04:43.11 	 & 25.19 	 & 0.08 	 & 24.14 	 & 0.15 \\
256 	 & 10:21:45.11 	 & +18:04:46.25 	 & 25.22 	 & 0.09 	 & 24.12 	 & 0.15 \\
257 	 & 10:21:45.30 	 & +18:06:00.66 	 & 25.24 	 & 0.09 	 & 24.08 	 & 0.15 \\
258 	 & 10:21:45.67 	 & +18:04:46.76 	 & 25.09 	 & 0.08 	 & 24.47 	 & 0.18 \\
259 	 & 10:21:44.59 	 & +18:05:35.56 	 & 25.20 	 & 0.09 	 & 24.17 	 & 0.16 \\
260 	 & 10:21:45.45 	 & +18:05:32.28 	 & 25.24 	 & 0.09 	 & 24.11 	 & 0.15 \\
261 	 & 10:21:43.80 	 & +18:05:48.56 	 & 25.64 	 & 0.12 	 & 23.92 	 & 0.14 \\
262 	 & 10:21:42.80 	 & +18:05:45.24 	 & 25.30 	 & 0.09 	 & 24.11 	 & 0.15 \\
263 	 & 10:21:43.62 	 & +18:05:53.35 	 & 25.18 	 & 0.08 	 & 24.34 	 & 0.17 \\
264 	 & 10:21:42.25 	 & +18:05:16.05 	 & 25.17 	 & 0.08 	 & 24.33 	 & 0.17 \\
265 	 & 10:21:45.75 	 & +18:05:02.20 	 & 25.27 	 & 0.09 	 & 24.22 	 & 0.16 \\
266 	 & 10:21:45.18 	 & +18:05:46.06 	 & 25.14 	 & 0.08 	 & 24.43 	 & 0.18 \\
267 	 & 10:21:44.56 	 & +18:05:40.03 	 & 25.29 	 & 0.09 	 & 24.16 	 & 0.16 \\
268 	 & 10:21:43.99 	 & +18:05:34.25 	 & 25.23 	 & 0.09 	 & 24.30 	 & 0.17 \\
269 	 & 10:21:43.26 	 & +18:06:04.59 	 & 25.34 	 & 0.09 	 & 24.14 	 & 0.15 \\
270 	 & 10:21:44.03 	 & +18:05:10.83 	 & 25.34 	 & 0.09 	 & 24.15 	 & 0.15 \\
271 	 & 10:21:43.95 	 & +18:05:07.22 	 & 25.46 	 & 0.10 	 & 24.09 	 & 0.15 \\
272 	 & 10:21:45.59 	 & +18:05:12.84 	 & 25.28 	 & 0.09 	 & 24.29 	 & 0.17 \\
273 	 & 10:21:43.32 	 & +18:05:55.21 	 & 25.19 	 & 0.08 	 & 24.49 	 & 0.18 \\
274 	 & 10:21:43.78 	 & +18:05:38.43 	 & 25.14 	 & 0.08 	 & 24.93 	 & 0.24 \\
275 	 & 10:21:44.06 	 & +18:06:05.82 	 & 25.27 	 & 0.09 	 & 24.36 	 & 0.17 \\
276 	 & 10:21:44.25 	 & +18:04:50.57 	 & 25.20 	 & 0.09 	 & 24.66 	 & 0.20 \\
277 	 & 10:21:44.25 	 & +18:05:32.00 	 & 25.25 	 & 0.09 	 & 24.47 	 & 0.18 \\
278 	 & 10:21:43.98 	 & +18:05:45.24 	 & 25.25 	 & 0.09 	 & 24.48 	 & 0.18 \\
279 	 & 10:21:45.47 	 & +18:04:54.71 	 & 25.25 	 & 0.09 	 & 24.47 	 & 0.18 \\
280 	 & 10:21:45.04 	 & +18:05:34.31 	 & 25.42 	 & 0.10 	 & 24.21 	 & 0.16 \\
281 	 & 10:21:45.18 	 & +18:04:56.16 	 & 25.29 	 & 0.09 	 & 24.42 	 & 0.18 \\
282 	 & 10:21:44.50 	 & +18:05:23.23 	 & 25.22 	 & 0.09 	 & 24.68 	 & 0.21 \\
283 	 & 10:21:45.22 	 & +18:05:20.87 	 & 25.42 	 & 0.10 	 & 24.29 	 & 0.17 \\
284 	 & 10:21:44.47 	 & +18:05:34.85 	 & 25.39 	 & 0.10 	 & 24.32 	 & 0.17 \\
285 	 & 10:21:42.80 	 & +18:05:02.94 	 & 25.43 	 & 0.10 	 & 24.25 	 & 0.16 \\
286 	 & 10:21:42.77 	 & +18:06:03.24 	 & 25.28 	 & 0.09 	 & 24.50 	 & 0.18 \\
287 	 & 10:21:45.58 	 & +18:05:07.51 	 & 25.31 	 & 0.09 	 & 24.42 	 & 0.18 \\
288 	 & 10:21:47.20 	 & +18:04:33.48 	 & 25.31 	 & 0.09 	 & 24.45 	 & 0.18 \\
289 	 & 10:21:44.07 	 & +18:06:02.35 	 & 25.21 	 & 0.09 	 & 24.96 	 & 0.25 \\
290 	 & 10:21:44.07 	 & +18:05:29.07 	 & 25.30 	 & 0.09 	 & 24.50 	 & 0.18 \\
291 	 & 10:21:45.02 	 & +18:05:47.85 	 & 25.52 	 & 0.11 	 & 24.18 	 & 0.16 \\
292 	 & 10:21:45.20 	 & +18:05:23.68 	 & 25.26 	 & 0.09 	 & 24.69 	 & 0.21 \\
293 	 & 10:21:45.54 	 & +18:04:28.21 	 & 25.23 	 & 0.09 	 & 24.87 	 & 0.23 \\
294 	 & 10:21:44.03 	 & +18:05:51.38 	 & 25.34 	 & 0.09 	 & 24.52 	 & 0.19 \\
295 	 & 10:21:41.72 	 & +18:06:12.70 	 & 25.59 	 & 0.11 	 & 24.23 	 & 0.16 \\
296 	 & 10:21:46.65 	 & +18:05:02.93 	 & 25.31 	 & 0.09 	 & 24.73 	 & 0.21 \\
297 	 & 10:21:43.50 	 & +18:06:08.94 	 & 25.40 	 & 0.10 	 & 24.44 	 & 0.18 \\
298 	 & 10:21:42.84 	 & +18:06:05.98 	 & 25.46 	 & 0.10 	 & 24.53 	 & 0.19 \\
299 	 & 10:21:43.78 	 & +18:06:08.06 	 & 25.52 	 & 0.11 	 & 24.32 	 & 0.17 \\
300 	 & 10:21:41.87 	 & +18:05:50.47 	 & 25.36 	 & 0.10 	 & 24.60 	 & 0.20 \\
301 	 & 10:21:44.17 	 & +18:05:42.81 	 & 25.53 	 & 0.11 	 & 24.32 	 & 0.17 \\
302 	 & 10:21:45.21 	 & +18:05:32.78 	 & 25.29 	 & 0.09 	 & 25.12 	 & 0.28 \\
303 	 & 10:21:46.02 	 & +18:05:06.51 	 & 25.42 	 & 0.10 	 & 24.54 	 & 0.19 \\
304 	 & 10:21:43.11 	 & +18:05:51.07 	 & 25.66 	 & 0.12 	 & 24.30 	 & 0.17 \\
305 	 & 10:21:44.25 	 & +18:05:35.65 	 & 25.41 	 & 0.10 	 & 24.50 	 & 0.18 \\
306 	 & 10:21:42.10 	 & +18:05:31.10 	 & 25.32 	 & 0.09 	 & 25.03 	 & 0.26 \\
307 	 & 10:21:44.33 	 & +18:05:11.19 	 & 25.62 	 & 0.12 	 & 24.64 	 & 0.20 \\
308 	 & 10:21:46.12 	 & +18:05:16.27 	 & 25.54 	 & 0.11 	 & 24.58 	 & 0.19 \\
309 	 & 10:21:41.89 	 & +18:05:43.21 	 & 25.75 	 & 0.13 	 & 24.29 	 & 0.16 \\
310 	 & 10:21:45.72 	 & +18:05:06.80 	 & 25.55 	 & 0.11 	 & 24.35 	 & 0.17 \\
311 	 & 10:21:45.37 	 & +18:05:50.72 	 & 25.52 	 & 0.11 	 & 24.44 	 & 0.18 \\
312 	 & 10:21:44.12 	 & +18:05:37.83 	 & 25.46 	 & 0.10 	 & 24.55 	 & 0.19 \\
313 	 & 10:21:43.52 	 & +18:06:01.43 	 & 25.65 	 & 0.12 	 & 24.31 	 & 0.17 \\
314 	 & 10:21:42.83 	 & +18:05:47.81 	 & 25.66 	 & 0.12 	 & 24.31 	 & 0.17 \\
315 	 & 10:21:45.91 	 & +18:05:04.61 	 & 25.37 	 & 0.10 	 & 25.07 	 & 0.27 \\
316 	 & 10:21:44.46 	 & +18:05:22.42 	 & 25.34 	 & 0.09 	 & 25.07 	 & 0.27 \\
317 	 & 10:21:44.86 	 & +18:05:44.65 	 & 25.56 	 & 0.11 	 & 24.45 	 & 0.18 \\
318 	 & 10:21:44.49 	 & +18:05:46.08 	 & 25.62 	 & 0.12 	 & 24.38 	 & 0.17 \\
319 	 & 10:21:44.13 	 & +18:06:15.03 	 & 25.46 	 & 0.10 	 & 24.70 	 & 0.21 \\
320 	 & 10:21:46.56 	 & +18:04:39.20 	 & 25.52 	 & 0.11 	 & 24.50 	 & 0.18 \\
321 	 & 10:21:45.25 	 & +18:05:10.43 	 & 25.41 	 & 0.10 	 & 24.92 	 & 0.24 \\
322 	 & 10:21:44.07 	 & +18:05:22.34 	 & 25.59 	 & 0.11 	 & 24.48 	 & 0.18 \\
323 	 & 10:21:44.56 	 & +18:06:03.50 	 & 25.52 	 & 0.11 	 & 24.55 	 & 0.19 \\
324 	 & 10:21:43.56 	 & +18:06:00.36 	 & 25.50 	 & 0.10 	 & 24.57 	 & 0.19 \\
325 	 & 10:21:45.01 	 & +18:04:37.07 	 & 25.58 	 & 0.11 	 & 24.47 	 & 0.18 \\
326 	 & 10:21:44.34 	 & +18:05:51.01 	 & 25.59 	 & 0.11 	 & 24.47 	 & 0.18 \\
327 	 & 10:21:43.15 	 & +18:04:51.15 	 & 25.37 	 & 0.10 	 & 24.94 	 & 0.24 \\
328 	 & 10:21:46.09 	 & +18:05:04.75 	 & 25.60 	 & 0.11 	 & 24.50 	 & 0.18 \\
329 	 & 10:21:42.40 	 & +18:05:58.11 	 & 25.50 	 & 0.10 	 & 24.68 	 & 0.21 \\
330 	 & 10:21:44.07 	 & +18:05:27.05 	 & 25.54 	 & 0.11 	 & 24.78 	 & 0.22 \\
331 	 & 10:21:45.07 	 & +18:05:36.05 	 & 25.70 	 & 0.12 	 & 24.42 	 & 0.18 \\
332 	 & 10:21:45.83 	 & +18:05:38.59 	 & 25.49 	 & 0.10 	 & 24.83 	 & 0.23 \\
333 	 & 10:21:43.32 	 & +18:05:19.56 	 & 25.64 	 & 0.12 	 & 24.55 	 & 0.19 \\
334 	 & 10:21:46.07 	 & +18:04:50.66 	 & 25.57 	 & 0.11 	 & 24.68 	 & 0.21 \\
335 	 & 10:21:43.43 	 & +18:06:25.72 	 & 25.95 	 & 0.15 	 & 24.35 	 & 0.17 \\
336 	 & 10:21:44.20 	 & +18:05:07.38 	 & 25.69 	 & 0.12 	 & 24.50 	 & 0.18 \\
337 	 & 10:21:44.10 	 & +18:05:13.34 	 & 25.56 	 & 0.11 	 & 24.73 	 & 0.21 \\
338 	 & 10:21:44.27 	 & +18:05:08.88 	 & 25.52 	 & 0.11 	 & 24.77 	 & 0.22 \\
339 	 & 10:21:46.51 	 & +18:05:28.99 	 & 25.50 	 & 0.10 	 & 24.91 	 & 0.24 \\
340 	 & 10:21:46.03 	 & +18:05:17.04 	 & 25.62 	 & 0.12 	 & 24.59 	 & 0.19 \\
341 	 & 10:21:42.01 	 & +18:06:01.25 	 & 25.51 	 & 0.10 	 & 24.96 	 & 0.25 \\
342 	 & 10:21:44.43 	 & +18:05:41.90 	 & 25.78 	 & 0.13 	 & 24.46 	 & 0.18 \\
343 	 & 10:21:42.12 	 & +18:05:29.49 	 & 25.76 	 & 0.13 	 & 24.48 	 & 0.18 \\
344 	 & 10:21:45.19 	 & +18:05:04.38 	 & 25.73 	 & 0.13 	 & 24.59 	 & 0.19 \\
345 	 & 10:21:42.69 	 & +18:05:51.58 	 & 25.63 	 & 0.12 	 & 24.70 	 & 0.21 \\
346 	 & 10:21:46.96 	 & +18:04:55.85 	 & 25.51 	 & 0.10 	 & 25.11 	 & 0.28 \\
347 	 & 10:21:43.77 	 & +18:05:58.77 	 & 25.67 	 & 0.12 	 & 24.58 	 & 0.19 \\
348 	 & 10:21:43.22 	 & +18:05:34.71 	 & 25.98 	 & 0.15 	 & 24.41 	 & 0.17 \\
349 	 & 10:21:44.96 	 & +18:05:46.34 	 & 25.62 	 & 0.12 	 & 24.68 	 & 0.21 \\
350 	 & 10:21:46.45 	 & +18:05:13.94 	 & 25.53 	 & 0.11 	 & 24.93 	 & 0.24 \\
351 	 & 10:21:45.55 	 & +18:04:55.37 	 & 25.76 	 & 0.13 	 & 24.52 	 & 0.19 \\
352 	 & 10:21:44.07 	 & +18:05:35.25 	 & 25.57 	 & 0.11 	 & 24.88 	 & 0.24 \\
353 	 & 10:21:43.43 	 & +18:05:33.54 	 & 25.56 	 & 0.11 	 & 25.12 	 & 0.28 \\
354 	 & 10:21:45.01 	 & +18:05:42.17 	 & 25.60 	 & 0.11 	 & 24.82 	 & 0.23 \\
355 	 & 10:21:45.19 	 & +18:04:57.59 	 & 25.58 	 & 0.11 	 & 24.94 	 & 0.25 \\
356 	 & 10:21:42.37 	 & +18:06:06.47 	 & 25.61 	 & 0.12 	 & 24.82 	 & 0.23 \\
357 	 & 10:21:47.25 	 & +18:04:48.19 	 & 25.60 	 & 0.11 	 & 25.11 	 & 0.28 \\
358 	 & 10:21:44.09 	 & +18:04:57.32 	 & 25.96 	 & 0.15 	 & 24.49 	 & 0.18 \\
359 	 & 10:21:41.72 	 & +18:05:55.85 	 & 25.69 	 & 0.12 	 & 24.83 	 & 0.23 \\
360 	 & 10:21:46.71 	 & +18:04:41.98 	 & 25.70 	 & 0.12 	 & 24.73 	 & 0.21 \\
361 	 & 10:21:43.56 	 & +18:04:42.74 	 & 25.75 	 & 0.13 	 & 24.65 	 & 0.20 \\
362 	 & 10:21:46.39 	 & +18:05:26.69 	 & 26.02 	 & 0.16 	 & 24.49 	 & 0.18 \\
363 	 & 10:21:43.06 	 & +18:06:05.00 	 & 25.82 	 & 0.14 	 & 24.65 	 & 0.20 \\
364 	 & 10:21:46.71 	 & +18:04:28.63 	 & 25.69 	 & 0.12 	 & 24.96 	 & 0.25 \\
365 	 & 10:21:46.58 	 & +18:04:56.71 	 & 25.70 	 & 0.12 	 & 24.91 	 & 0.24 \\
366 	 & 10:21:45.09 	 & +18:04:53.31 	 & 25.72 	 & 0.12 	 & 25.13 	 & 0.28 \\
367 	 & 10:21:44.60 	 & +18:05:37.37 	 & 25.69 	 & 0.12 	 & 24.93 	 & 0.24 \\
368 	 & 10:21:43.19 	 & +18:06:09.29 	 & 25.75 	 & 0.13 	 & 24.77 	 & 0.22 \\
369 	 & 10:21:43.54 	 & +18:05:44.43 	 & 25.77 	 & 0.13 	 & 24.94 	 & 0.25 \\
370 	 & 10:21:42.17 	 & +18:06:03.30 	 & 25.71 	 & 0.12 	 & 24.92 	 & 0.24 \\
371 	 & 10:21:43.71 	 & +18:06:00.39 	 & 25.78 	 & 0.13 	 & 24.75 	 & 0.21 \\
372 	 & 10:21:46.04 	 & +18:05:02.46 	 & 25.83 	 & 0.14 	 & 24.69 	 & 0.21 \\
373 	 & 10:21:45.49 	 & +18:05:06.41 	 & 25.74 	 & 0.13 	 & 24.86 	 & 0.23 \\
374 	 & 10:21:46.61 	 & +18:05:05.19 	 & 25.90 	 & 0.14 	 & 24.62 	 & 0.20 \\
375 	 & 10:21:43.24 	 & +18:05:51.32 	 & 26.14 	 & 0.18 	 & 24.67 	 & 0.20 \\
376 	 & 10:21:42.62 	 & +18:06:00.26 	 & 25.74 	 & 0.13 	 & 25.09 	 & 0.28 \\
377 	 & 10:21:43.55 	 & +18:05:18.01 	 & 26.08 	 & 0.17 	 & 24.62 	 & 0.20 \\
378 	 & 10:21:45.09 	 & +18:05:57.25 	 & 25.95 	 & 0.15 	 & 24.78 	 & 0.22 \\
379 	 & 10:21:42.04 	 & +18:05:46.70 	 & 25.83 	 & 0.14 	 & 24.72 	 & 0.25 \\
380 	 & 10:21:45.77 	 & +18:04:34.49 	 & 25.95 	 & 0.15 	 & 24.73 	 & 0.21 \\
381 	 & 10:21:46.16 	 & +18:04:20.62 	 & 25.81 	 & 0.14 	 & 24.84 	 & 0.23 \\
382 	 & 10:21:42.38 	 & +18:05:31.17 	 & 25.96 	 & 0.15 	 & 24.71 	 & 0.21 \\
383 	 & 10:21:43.52 	 & +18:05:48.19 	 & 25.74 	 & 0.13 	 & 25.12 	 & 0.28 \\
384 	 & 10:21:46.33 	 & +18:05:07.99 	 & 25.84 	 & 0.14 	 & 24.88 	 & 0.24 \\
385 	 & 10:21:43.63 	 & +18:05:06.06 	 & 25.86 	 & 0.14 	 & 24.82 	 & 0.23 \\
386 	 & 10:21:43.67 	 & +18:06:06.26 	 & 25.75 	 & 0.13 	 & 25.07 	 & 0.27 \\
387 	 & 10:21:42.25 	 & +18:05:29.67 	 & 26.16 	 & 0.18 	 & 24.66 	 & 0.20 \\
388 	 & 10:21:43.54 	 & +18:06:21.91 	 & 25.80 	 & 0.13 	 & 25.02 	 & 0.26 \\
389 	 & 10:21:43.14 	 & +18:05:49.84 	 & 25.87 	 & 0.14 	 & 24.86 	 & 0.23 \\
390 	 & 10:21:42.71 	 & +18:05:56.01 	 & 26.08 	 & 0.17 	 & 24.67 	 & 0.20 \\
391 	 & 10:21:43.38 	 & +18:05:57.36 	 & 25.80 	 & 0.13 	 & 25.09 	 & 0.28 \\
392 	 & 10:21:46.77 	 & +18:04:22.61 	 & 26.02 	 & 0.16 	 & 24.73 	 & 0.21 \\
393 	 & 10:21:44.07 	 & +18:05:36.09 	 & 26.00 	 & 0.16 	 & 24.98 	 & 0.25 \\
394 	 & 10:21:45.97 	 & +18:04:40.93 	 & 26.04 	 & 0.16 	 & 25.02 	 & 0.26 \\
395 	 & 10:21:46.15 	 & +18:05:11.25 	 & 26.07 	 & 0.17 	 & 24.73 	 & 0.21 \\
396 	 & 10:21:43.65 	 & +18:05:34.37 	 & 26.29 	 & 0.20 	 & 24.68 	 & 0.21 \\
397 	 & 10:21:43.75 	 & +18:05:37.73 	 & 25.91 	 & 0.15 	 & 24.89 	 & 0.24 \\
398 	 & 10:21:46.80 	 & +18:04:48.77 	 & 25.84 	 & 0.14 	 & 25.00 	 & 0.26 \\
399 	 & 10:21:43.81 	 & +18:05:04.15 	 & 25.91 	 & 0.15 	 & 24.94 	 & 0.25 \\
400 	 & 10:21:45.85 	 & +18:04:25.14 	 & 26.00 	 & 0.16 	 & 24.88 	 & 0.24 \\
401 	 & 10:21:45.16 	 & +18:05:57.98 	 & 25.87 	 & 0.14 	 & 25.07 	 & 0.27 \\
402 	 & 10:21:43.55 	 & +18:05:22.20 	 & 26.46 	 & 0.23 	 & 24.71 	 & 0.21 \\
403 	 & 10:21:45.92 	 & +18:05:54.19 	 & 25.90 	 & 0.14 	 & 24.96 	 & 0.25 \\
404 	 & 10:21:44.62 	 & +18:04:27.49 	 & 26.11 	 & 0.17 	 & 24.84 	 & 0.23 \\
405 	 & 10:21:44.35 	 & +18:05:40.35 	 & 26.00 	 & 0.16 	 & 24.84 	 & 0.23 \\
406 	 & 10:21:42.86 	 & +18:06:00.76 	 & 26.05 	 & 0.17 	 & 24.78 	 & 0.22 \\
407 	 & 10:21:42.69 	 & +18:06:09.39 	 & 25.95 	 & 0.15 	 & 25.10 	 & 0.28 \\
408 	 & 10:21:47.08 	 & +18:04:59.24 	 & 26.35 	 & 0.21 	 & 24.80 	 & 0.22 \\
409 	 & 10:21:44.24 	 & +18:05:56.05 	 & 26.02 	 & 0.16 	 & 24.94 	 & 0.24 \\
410 	 & 10:21:43.68 	 & +18:05:56.25 	 & 26.05 	 & 0.17 	 & 25.04 	 & 0.27 \\
411 	 & 10:21:42.77 	 & +18:05:42.05 	 & 26.31 	 & 0.21 	 & 24.74 	 & 0.21 \\
412 	 & 10:21:43.30 	 & +18:06:10.34 	 & 26.04 	 & 0.16 	 & 25.00 	 & 0.26 \\
413 	 & 10:21:45.71 	 & +18:04:51.56 	 & 26.04 	 & 0.16 	 & 25.01 	 & 0.26 \\
414 	 & 10:21:43.13 	 & +18:05:51.97 	 & 26.12 	 & 0.17 	 & 24.88 	 & 0.23 \\
415 	 & 10:21:43.20 	 & +18:06:10.58 	 & 26.19 	 & 0.19 	 & 24.88 	 & 0.23 \\
416 	 & 10:21:46.61 	 & +18:04:29.66 	 & 26.03 	 & 0.16 	 & 24.98 	 & 0.25 \\
417 	 & 10:21:44.18 	 & +18:06:00.51 	 & 26.21 	 & 0.19 	 & 24.88 	 & 0.23 \\
418 	 & 10:21:42.70 	 & +18:06:24.60 	 & 26.27 	 & 0.20 	 & 24.84 	 & 0.23 \\
419 	 & 10:21:42.83 	 & +18:06:13.82 	 & 26.05 	 & 0.17 	 & 25.07 	 & 0.27 \\
420 	 & 10:21:44.96 	 & +18:04:48.99 	 & 26.04 	 & 0.16 	 & 25.14 	 & 0.28 \\
421 	 & 10:21:44.60 	 & +18:06:18.84 	 & 26.25 	 & 0.20 	 & 24.91 	 & 0.24 \\
422 	 & 10:21:44.37 	 & +18:05:53.14 	 & 26.49 	 & 0.24 	 & 24.98 	 & 0.25 \\
423 	 & 10:21:46.13 	 & +18:04:19.59 	 & 26.14 	 & 0.18 	 & 25.02 	 & 0.26 \\
424 	 & 10:21:42.86 	 & +18:05:59.05 	 & 26.11 	 & 0.17 	 & 25.08 	 & 0.27 \\
425 	 & 10:21:47.44 	 & +18:04:56.36 	 & 26.30 	 & 0.20 	 & 24.92 	 & 0.24 \\
426 	 & 10:21:42.13 	 & +18:05:21.43 	 & 26.41 	 & 0.23 	 & 24.93 	 & 0.24 \\
427 	 & 10:21:44.62 	 & +18:05:58.95 	 & 26.30 	 & 0.20 	 & 25.00 	 & 0.26 \\
428 	 & 10:21:44.45 	 & +18:05:38.33 	 & 26.62 	 & 0.27 	 & 24.91 	 & 0.24 \\
429 	 & 10:21:45.61 	 & +18:05:31.33 	 & 26.39 	 & 0.22 	 & 24.98 	 & 0.25 \\
430 	 & 10:21:42.66 	 & +18:05:09.74 	 & 26.32 	 & 0.21 	 & 25.05 	 & 0.27 \\
431 	 & 10:21:46.18 	 & +18:04:40.96 	 & 26.34 	 & 0.21 	 & 25.07 	 & 0.27 \\
432 	 & 10:21:42.34 	 & +18:05:25.39 	 & 26.32 	 & 0.21 	 & 25.15 	 & 0.29 \\
433 	 & 10:21:45.23 	 & +18:04:47.46 	 & 26.42 	 & 0.23 	 & 25.07 	 & 0.27 \\
434 	 & 10:21:47.30 	 & +18:04:38.55 	 & 26.33 	 & 0.21 	 & 24.88 	 & 0.28 \\
435 	 & 10:21:42.70 	 & +18:05:48.63 	 & 26.30 	 & 0.20 	 & 25.13 	 & 0.28 \\
436 	 & 10:21:45.74 	 & +18:04:47.50 	 & 26.48 	 & 0.24 	 & 25.12 	 & 0.28 \\
437 	 & 10:21:43.19 	 & +18:06:29.38 	 & 26.33 	 & 0.21 	 & 25.18 	 & 0.29 \\
438 	 & 10:21:44.43 	 & +18:06:02.87 	 & 26.45 	 & 0.23 	 & 25.13 	 & 0.28 \\
\label{tab:catalogue}
\end{longtable}}
\end{center}

\clearpage
\begin{table}
\begin{center}
\caption{Comparison of Physical Characteristics}
\label{tab:parameters}
\end{center}
\begin{center}
\begin{tabular}{lccc}
\hline 
\hline 
	& M$_V$	& M$_{stars}$		& $\mu_V$		\\
Galaxy	& (mag)	& ($\times 10^5$ \msun)	& (mag arcsec$^{-2}$)	\\
\hline
Leo~P	& -9.4	& 5.7			& 24.5$\pm0.6$	    	\\
Carina	& -9.1	& 3.8			& 25.5$\pm0.5$		\\
Sextans	& -9.3	& 4.4			& 27.1$\pm0.5$		\\
Leo~II	& -9.8	& 7.4			& 24.2$\pm0.6$		\\

\hline
 \\
\end{tabular}
\end{center}
\tablecomments{The physical properties of Leo~P are most similar to those of the more massive dSphs in the LG such as Carina, Sextans, and Leo~II \citep{Irwin1995}.}
\end{table}


\begin{thebibliography}{}
\bibitem[Aloisi et al.(2007)]{Aloisi2007} Aloisi, A., Clementini, 
G., Tosi, M., et al.\ 2007, \apjl, 667, L151 
\bibitem[Asplund et al.(2009)]{Asplund2009} Asplund, M., Grevesse, N., Sauval, A.~J., \& Scott, P.\ 2009, \araa, 47, 481 
\bibitem[Bell \& de Jong(2001)]{Bell2001} Bell, E.~F., \& de Jong, R.~S.\ 2001, \apj, 550, 212 
\bibitem[Berg et al.(2012)]{Berg2012} Berg, D.~A., Skillman, 
E.~D., Marble, A.~R., et al.\ 2012, \apj, 754, 98 
\bibitem[Bessell(1979)]{Bessel1979} Bessell, M.~S.\ 1979, \pasp, 
91, 589 
\bibitem[Bernstein-Cooper et al.(2013)]{Bernstein-Cooper2013} Bernstein-Cooper, E. et al.\ in prep
\bibitem[Cole et al.(2007)]{Cole2007} Cole, A.~A., Skillman, 
E.~D., Tolstoy, E., et al.\ 2007, \apjl, 659, L17 
\bibitem[Binggeli et al.(1988)]{Binggeli1988} Binggeli, B., Sandage, A., \& Tammann, G.~A.\ 1988, \araa, 26, 509 
\bibitem[Cox(2000)]{Cox2000} Cox, A.~N.\ 2000, Allen's 
Astrophysical Quantities,  
\bibitem[Da Costa \& Armandroff(1990)]{DaCosta1990} Da Costa, G.~S., \& Armandroff, T.~E.\ 1990, \aj, 100, 162 
\bibitem[Dalcanton et al.(2009)]{Dalcanton2009} Dalcanton, J.~J., 
Williams, B.~F., Seth, A.~C., et al.\ 2009, \apjs, 183, 67 
\bibitem[de Vaucouleurs(1953)]{deVaucouleurs1953} de Vaucouleurs, G.\ 
1953, \aj, 58, 30 
\bibitem[Dekel \& Silk(1986)]{Dekel1986} Dekel, A., \& Silk, J.\ 1986, \apj, 303, 39 
\bibitem[Dolphin(2000)]{Dolphin2000} Dolphin, A.~E.\ 2000, \pasp, 
112, 1383 
\bibitem[Dolphin(2002)]{Dolphin2002} Dolphin, A.~E.\ 2002, \mnras, 
332, 91 
\bibitem[Ekta et al.(2006)]{Ekta2006} Ekta, Chengalur, J.~N., 
\& Pustilnik, S.~A.\ 2006, \mnras, 372, 853 
\bibitem[Ekta et al.(2008)]{Ekta2008} Ekta, Chengalur, J.~N., 
\& Pustilnik, S.~A.\ 2008, \mnras, 391, 881 
\bibitem[Ekta et al.(2009)]{Ekta2009} Ekta, B., Pustilnik, 
S.~A., \& Chengalur, J.~N.\ 2009, \mnras, 397, 963 
\bibitem[Ekta \& Chengalur(2010a)]{Ekta2010a} Ekta, B., \& Chengalur, J.~N.\ 2010, \mnras, 403, 295 
\bibitem[Ekta 
\& Chengalur(2010b)]{Ekta2010b} Ekta, B., \& Chengalur, J.~N.\ 2010, \mnras, 406, 1238 
\bibitem[Faber \& Lin(1983)]{Faber1983} Faber, S.~M., \& Lin, D.~N.~C.\ 1983, \apjl, 266, L17 
\bibitem[Freedman(1988)]{Freedman1988} Freedman, W.~L.\ 1988, \apj, 
326, 691 
\bibitem[Giovanelli et al.(2005)]{Giovanelli2005} Giovanelli, R., 
Haynes, M.~P., Kent, B.~R., et al.\ 2005, \aj, 130, 2598 
\bibitem[Giovanelli et al.(2013)]{Giovanelli2013} Giovanelli, R., 
Haynes, M.~P., Adams, E.~A.~K., et al.\ 2013, \aj, 146, 15 
\bibitem[Girardi \& Marigo(2007)]{Girardi2007} Girardi, L., \& Marigo, P.\ 2007, \aap, 462, 237 
\bibitem[Girardi et al.(2010)]{Girardi2010} Girardi, L., Williams, 
B.~F., Gilbert, K.~M., et al.\ 2010, \apj, 724, 1030 
\bibitem[Grebel et al.(2003)]{Grebel2003} Grebel, E.~K., 
Gallagher, J.~S., III, \& Harbeck, D.\ 2003, \aj, 125, 1926 
\bibitem[Haynes et al.(2011)]{Haynes2011} Haynes, M.~P., 
Giovanelli, R., Martin, A.~M., et al.\ 2011, \aj, 142, 170 
\bibitem[Irwin \& Hatzidimitriou(1995)]{Irwin1995} Irwin, M., \& Hatzidimitriou, D.\ 1995, \mnras, 277, 1354 
\bibitem[Karachentsev et al.(2009)]{Karachentsev2009} Karachentsev, 
I.~D., Kashibadze, O.~G., Makarov, D.~I., 
\& Tully, R.~B.\ 2009, \mnras, 393, 1265 
\bibitem[Kormendy(1985)]{Kormendy1985} Kormendy, J.\ 1985, \apj, 
295, 73 
\bibitem[Kunth \& \"{O}stlin(2000)]{Kunth2000} Kunth, D., \& \"{O}stlin, G.\ 2000, \aapr, 10, 1 
\bibitem[Izotov \& Thuan(1999)]{Izotov1999} Izotov, Y.~I., \& Thuan, T.~X.\ 1999, \apj, 511, 639 
\bibitem[Izotov et al.(2005)]{Izotov2005} Izotov, Y.~I., Thuan, 
T.~X., \& Guseva, N.~G.\ 2005, \apj, 632, 210 
\bibitem[Izotov \& Thuan(2007)]{Izotov2007} Izotov, Y.~I., \& Thuan, T.~X.\ 2007, \apj, 665, 1115 
\bibitem[Landolt(1992)]{Landolt1992} Landolt, A.~U.\ 1992, \aj, 
104, 340 
\bibitem[Lebouteiller et al.(2013)]{Lebouteiller2013} Lebouteiller, V., 
Heap, S., Hubeny, I., \& Kunth, D.\ 2013, arXiv:1302.4746 
\bibitem[Lee et al.(1993)]{Lee1993} Lee, M.~G., Freedman, 
W.~L., \& Madore, B.~F.\ 1993, \apj, 417, 553 
\bibitem[Madore \& Freedman(1998)]{Madore1998} Madore, B.~F., \& Freedman, W.~L.\ 1998, Stellar astrophysics for the local group: VIII Canary Islands Winter School of Astrophysics, 263 
\bibitem[Makarov et al.(2006)]{Makarov2006} Makarov, D., Makarova, 
L., Rizzi, L., et al.\ 2006, \aj, 132, 2729 
\bibitem[Marigo et al.(2008)]{Marigo2008} Marigo, P., Girardi, L., Bressan, A., et al.\ 2008, \aap, 482, 883 
\bibitem[Mateo(1998)]{Mateo1998} Mateo, M.~L.\ 1998, \araa, 36, 435 
\bibitem[Mayer et al.(2001a)]{Mayer2001a} Mayer, L., Governato, F., 
Colpi, M., et al.\ 2001, \apj, 559, 754 
\bibitem[Mayer et al.(2001b)]{Mayer2001b} Mayer, L., Governato, F., 
Colpi, M., et al.\ 2001, \apjl, 547, L123 
\bibitem[Mayer et al.(2006)]{Mayer2006} Mayer, L., Mastropietro, 
C., Wadsley, J., Stadel, J., \& Moore, B.\ 2006, \mnras, 369, 1021 
\bibitem[McConnachie(2012)]{McConnachie2012} McConnachie, A.~W.\ 2012, 
\aj, 144, 4 
\bibitem[McQuinn et al.(2010)]{McQuinn2010} McQuinn, K.~B.~W., 
Skillman, E.~D., Cannon, J.~M., et al.\ 2010, \apj, 721, 297 
\bibitem[Mould \& Kristian(1986)]{Mould1986} Mould, J., \& Kristian, J.\ 1986, \apj, 305, 591 
\bibitem[Pustilnik et al.(2005)]{Pustilnik2005} Pustilnik, S.~A., Kniazev, A.~Y., \& Pramskij, A.~G.\ 2005, \aap, 443, 91 
\bibitem[Pustilnik et al.(2004)]{Pustilnik2004} Pustilnik, S.~A., Pramskij, A.~G., \& Kniazev, A.~Y.\ 2004, \aap, 425, 51 
\bibitem[Rhode et al.(2013)]{Rhode2013} Rhode, K.~L., Salzer, 
J.~J., Haurberg, N.~C., et al.\ 2013, \aj, 145, 149 
\bibitem[Rizzi et al.(2007)]{Rizzi2007} Rizzi, L., Tully, R.~B., 
Makarov, D., et al.\ 2007, \apj, 661, 815 
\bibitem[Robin et al.(2003)]{Robin2003} Robin, A.~C., Reyl{\'e}, C., Derri{\`e}re, S., \& Picaud, S.\ 2003, \aap, 409, 523 
\bibitem[Sakai et al.(1996)]{Sakai1996} Sakai, S., Madore, B.~F., 
\& Freedman, W.~L.\ 1996, \apj, 461, 713 
\bibitem[Sakai et al.(1997)]{Sakai1997} Sakai, S., Madore, B.~F., 
\& Freedman, W.~L.\ 1997, \apj, 480, 589 
\bibitem[Schechter(1976)]{Schechter1976} Schechter, P.\ 1976, \apj, 
203, 297 
\bibitem[Schlafly \& Finkbeiner(2011)]{Schlafly2011} Schlafly, E.~F., \& Finkbeiner, D.~P.\ 2011, \apj, 737, 103 
\bibitem[Schlegel et al.(1998)]{Schlegel1998} Schlegel, D.~J., 
Finkbeiner, D.~P., \& Davis, M.\ 1998, \apj, 500, 525 
\bibitem[Skillman \& Kennicutt(1993)]{Skillman1993} Skillman, E.~D., \& Kennicutt, R.~C., Jr.\ 1993, \apj, 411, 655 
\bibitem[Skillman et al.(2013)]{Skillman2013} Skillman, E.~D., 
Salzer, J.~J., Berg, D.~A., et al.\ 2013, \aj, 146, 3 
\bibitem[Tolstoy et al.(2009)]{Tolstoy2009} Tolstoy, E., Hill, V., \& Tosi, M.\ 2009, \araa, 47, 371 
\bibitem[Tully et al.(2002)]{Tully2002} Tully, R.~B., Somerville, 
R.~S., Trentham, N., \& Verheijen, M.~A.~W.\ 2002, \apj, 569, 573 
\bibitem[Tully et al.(2006)]{Tully2006} Tully, R.~B., Rizzi, L., 
Dolphin, A.~E., et al.\ 2006, \aj, 132, 729 
\bibitem[Weisz et al.(2011)]{Weisz2011} Weisz, D.~R., Dalcanton, 
J.~J., Williams, B.~F., et al.\ 2011, \apj, 739, 5 
\end{thebibliography}
\end{document}